\documentclass[journal,onecolumn,draftcls]{IEEEtran}

\usepackage{cite}

\ifCLASSINFOpdf
\usepackage[pdftex]{graphicx}
\graphicspath{{figs/}}
\DeclareGraphicsExtensions{.pdf,.jpeg,.png,.jpg}
\else
\fi

\usepackage[cmex10]{amsmath}

\usepackage{array}

\usepackage{amsmath}
\usepackage{amssymb}

\usepackage{graphicx}

\usepackage{epstopdf}
\usepackage{float}
\usepackage{hyperref}

\usepackage[normalem]{ulem}

\input{MyMacros.sty}

\begin{document}

	\title{Information-Theoretic Privacy\\with General Distortion Constraints}
	\author{Kousha Kalantari~\IEEEmembership{Student Member,~IEEE}, 
		Oliver Kosut~\IEEEmembership{Member,~IEEE},
		Lalitha Sankar~\IEEEmembership{Senior Member,~IEEE}
		\thanks{Manuscript received August 18, 2017; This paper was presented in part at ISIT 2017~\cite{KoushaISIT2017}. This work is supported in part by the National Science Foundation under grant CCF-1422358.}%
		\thanks{The authors are with the  School of Electrical, Computer, and Energy Engineering, Arizona State University, Tempe, Arizona 85287, USA (e-mail: kousha.kalantari@asu.edu, okosut@asu.edu, lalithasankar@asu.edu).}
	}

	\maketitle
	\begin{abstract}
		Thee privacy-utility tradeoff problem is formulated as determining the privacy mechanism (random mapping) that minimizes the mutual information (a metric for privacy leakage) between the private features of the original dataset and a released version.
		The minimization is studied with two types of constraints on the distortion between the public features and the released version of the dataset: (i) subject to a constraint on the expected value of a cost function $f$ applied to the distortion, and (ii) subject to bounding the complementary CDF of the distortion by a non-increasing function $g$. The first scenario captures various practical cost functions for distorted released data, while the second scenario covers large deviation constraints on utility.
		The asymptotic optimal leakage is derived in both scenarios. For the distortion cost constraint, it is shown that for convex cost functions there is no asymptotic loss in using stationary memoryless mechanisms. For the complementary CDF bound on distortion, the asymptotic leakage is derived for general mechanisms and shown to be the integral of the single letter leakage function with respect to the Lebesgue---Stieltjes measure defined based on the refined bound on distortion. However, it is shown that memoryless mechanisms are generally suboptimal in both cases.
	\end{abstract}
	
	
	
	\section{Introduction}
	Let $(X^n,Y^n)$ be a random data sequence where $X$ and $Y$ represent the public and private sections of the data respectively, and are drawn from an i.i.d. distribution $P_{X,Y}$. Each entry $(X_i,Y_i)$ represents a row of the dataset. We wish to find a privacy mechanism, i.e. a random mapping, that reveals a sequence $\hat{X}^n$ such that (i) statistical information about $X^n$ can be learned from $\hat{X}^n$, and (ii) as little information as possible about private data $Y^n$ should be revealed by $\hat{X}^n$. These two goals are in conflict, since typically $X$ and $Y$ are correlated (especially when $X=Y$). Thus, we wish to characterize the privacy utility tradeoff (PUT) while being careful to choose meaningful utility and privacy metrics.
	
	Our focus is on inferential adversaries that can learn the hidden features $Y^n$ from the released dataset $\hX^n$. To this end, we motivate the use of mutual information between the private features $Y^n$ and the revealed version of the dataset $\hX^n$ as a metric for privacy leakage. Mutual information as a measure captures an adversary that refines its posterior belief of the private data from the released data, i.e., the adversary's loss function is the log-loss function \cite{CalmonFawaz2013}. Indeed other measures such as maximal leakage \cite{IssaTIT2020}, maximal correlation \cite{Calmon2013}, maximal $\alpha$-leakage \cite{LiaoTIT2019}, gain-function leakage \cite{KurriSK2024}, maximal $(\alpha,\beta)$-leakage \cite{GilaniKKS2024}, and pointwise information-density based measures \cite{SaeidianGSSO2026,TaylorVC2026} capture different adversarial goals, including guessing functions of the private data or bounding realization-dependent leakage. However, we restrict our focus here to a belief-refining adversary.

	For the choice of utility metric, the average distortion constraint in the form of $\mathbb{E}[d(X^n,\hX^n)] \le D$ has been used in many works, where $D$ is a distortion threshold and $d(\cdot,\cdot)$ is a given distortion function between public data and released data. However, this utility metric does not capture all aspects of distortion distribution.
	One possible step in order to capture more aspects of the distortion distribution, is via the \textit{tail probability constraint} (or equivalently called \textit{excess distortion constraint}). This has been of much interest in source coding (see for example \cite{Yeung,Kostina,Kostina2012c,YannisVerdu,Shkel}), channel coding (see for example \cite{YucelISIT2015,YucelAllerton2015,YucelISIT2016}) and studied in the context of privacy in \cite{KKSAllerton2016}. For a more detailed survey on finite blocklength approaches see \cite{tan2014asymptotic}.
	
	However, even the tail probability constraint does not capture the full spectrum of possibilities on applying bounds on distortion distribution. In this paper, we generalize the tail probability constraint in two ways:
	\begin{itemize}
		\item A bound $t$ on the average distortion cost, where the distortion cost is a non-decreasing function $f$ applied on a separable distortion measure $d$ between $X^n$ and $\hat{X}^n$. The resulting PUT is given by
		\begin{equation}\label{eq:optimization} 
		\begin{array}{ll}
		{\substack{\text{\normalsize{minimize}}\\ P_{\hX^n|X^n,Y^n} }} & \displaystyle  \frac{1}{n}I(Y^n;\hat{X}^n)\\
		\text{subject to} & \mathbb{E}[f(d(X^n,\hat{X}^n))]\le t,
		\end{array}
		\end{equation}
		
		\item A non-increasing function $g$ to bound the complementary CDF of the distortion measure $d$ between $X^n$ and $\hat{X}^n$. The resulting PUT is given by
		\begin{equation}\label{eq:optimization 2} 
		\begin{array}{ll}
		{\substack{\text{\normalsize{minimize}}\\ P_{\hX^n|X^n,Y^n} }} & \displaystyle  \frac{1}{n}I(Y^n;\hat{X}^n)\\
		\text{subject to} &  \mathbb{P}[d(X^n,\hX^n) > D] \le g(D), \forall D.
		\end{array}
		\end{equation}
	\end{itemize}
	The cost constraint in \eqref{eq:optimization} imposes increasing penalties on higher levels of distortion in general, and reduces to a tail probability constraint when $f(D)=\boldsymbol{1}(D> D_0)$, for some constant $D_0$. The distortion distribution bound in \eqref{eq:optimization 2} allows arbitrarily fine-tuned bounds on the complementary CDF of the distortion, and reduces to a tail probability constraint when $g(D)= 1 - (1-\epsilon) \boldsymbol{1}(D\ge D_0)$, for some constant $D_0$. Note that these two types of constraint are not equivalent in general and can capture different requirements on the distortion distribution.
	\subsection{Contributions}
	A privacy mechanism could be applied to a dataset as a whole, or to each individual entry of the dataset independently. We label the mechanisms for the two approaches as \textit{general} and \textit{memoryless} mechanisms. In this paper:
		\begin{itemize}
			\item We derive the asymptotic leakage distortion-cost tradeoff in \eqref{eq:optimization}. For memoryless mechanisms, it is equal to the single letter leakage function evaluated at the generalized inverse of the cost function whenever that inverse is uniquely defined, and at the countable set of discontinuity thresholds we provide matching single-letter upper and lower bounds. For general mechanisms, the asymptotic leakage is the lower convex envelope of the memoryless tradeoff curve.
			\item We also give the exact formulation of the asymptotic leakage in \eqref{eq:optimization 2} for memoryless and general mechanisms. For memoryless mechanisms, it is equal to the single letter leakage function evaluated at the largest distortion value at which $g(D)$ is equal to $1$. For general mechanisms, it is the integral of the single letter leakage function with respect to the Lebesgue---Stieltjes measure defined by the constraint function $g$.
			\item In both cases, the optimal general mechanisms are mixtures of memoryless mechanisms.
		\end{itemize} 
	The formulations in \eqref{eq:optimization} and \eqref{eq:optimization 2} include the dependence on both the public and private aspects of the dataset. In cases where the private data is not directly available, but the statistics are known, the private ($Y^n$), public ($X^n$), and revealed data ($\hX^n$) form a Markov chain $Y^n \rightarrow X^n \rightarrow \hX^n$. In this paper, we focus on the general case with both public and private data being available to the mechanism, but the results here generalize in a straightforward manner to the case when private data is not available.
	The post-2018 literature has substantially expanded the class of leakage measures and the design of privacy mechanisms, but the asymptotic role of non-linear distortion costs and full distortion-tail constraints remains distinct from these developments. Our contribution is therefore complementary: we keep the leakage measure fixed to mutual information and characterize how the entire distortion constraint, rather than only the single-letter average distortion threshold, changes the optimal first-order leakage and the need for general mechanisms.
	
	\subsection{Related Work}
	An alternative approach to more general distortion constraints is considered in \cite{Shkel} and referred to as \textit{$\tilde{f}$-separable distortion measures} \setcounter{footnote}{0} \footnote{We have changed their notation from $f$-separable to $\tilde{f}$-separable, in order to avoid confusion with our notation.}.
	In \cite{Shkel}, a multi-letter distortion measure $\tilde{d} (\cdot,\cdot)$ is defined as $\tilde{f}$-separable if
	\begin{equation}
	\tilde{d}(x^n, \hx^n) = \tilde{f}^{-1}\left(\frac{1}{n} \sum_{i=1}^{n} \tilde{f}(\tilde{d}(x_i,\hx_i))\right),
	\end{equation}
	for an increasing function $\tilde{f}$. The distortion cost constraints that we consider are more general in the sense that our notion of cost function $f$ applied to the distortion measure $d(\cdot,\cdot)$ covers a broader class of distortion constraints than an average bound on $\tilde{f}$-separable distortion measures studied in \cite{Shkel}. Specifically, the average constraint on an $\tilde{f}$-separable distortion measure has the form
	\begin{equation}
	\mathbb{E}\left[\tilde{f}^{-1} \left( \frac{1}{n} \sum_{i=1}^{n} \tilde{f}(\tilde{d}(x_i,\hx_i)) \right)\right] \le D,
	\end{equation}
	which clearly is a specific case for our formulation in \eqref{eq:optimization} that results from choosing $f= \tilde{f}^{-1}$ and $d(x,\hx) = \tilde{f}(\tilde{d}(x,\hx))$, such that $d(x^n,\hx^n) = \frac{1}{n} \sum_{i=1}^{n} d(x_i,\hx_i)$. Moreover, we allow for non-decreasing functions $f$, which means that $\tilde{f}$ does not have to be strictly increasing. We also note that our focus is on privacy rather than source coding.
	
	In the context of privacy, the privacy utility tradeoff with distinct $X$ and $Y$ is studied in \cite{Yamamoto} and more extensively in \cite{Sankar_TIFS_2013}, but the utility metric is only restricted to identity cost functions, i.e. $f(D)=D$. Generalizing this to the excess distortion constraint was considered by \cite{KKSAllerton2016}. In \cite{KKSAllerton2016}, we also differentiated between explicit availability or unavailability of the private data $Y$ to the privacy mechanism. Information theoretic approaches to privacy that are agnostic to the length of the dataset are considered in \cite{CalmonFawaz2013,CalmonMM15,Asoodeh2015}.
	
	Several recent works study neighboring privacy-utility tradeoffs with different utility or leakage models. Estimation-theoretic formulations quantify utility and privacy through guessing probability, mean-squared error, or chi-square information \cite{AsoodehDAL2019,WangVCMDV2019}. Other works replace mutual-information leakage by total variation, strong $\chi^2$ or $\ell_1$ criteria, or non-zero/per-letter privacy constraints \cite{RassouliGunduz2020TV,ZamaniOS2021,ZamaniOS2022,ZamaniOS2024}. Perfect-privacy and privacy-funnel variants ask how much information about useful data can be revealed under zero or bounded leakage \cite{RassouliGunduz2021Perfect,SreekumarGunduz2019}. Robustness of information-theoretic mechanisms to empirical distribution mismatch was studied in \cite{DiazWCS2020}. These papers are closest in modeling spirit to ours, but they do not characterize the first-order leakage under general multi-letter distortion-cost functions or under a prescribed complementary CDF bound on distortion.

	The hard-distortion formulation in \cite{LiaoKSC2018Hard} is also closely related because it provides deterministic fidelity guarantees under maximal $\alpha$-leakage. Our complementary CDF formulation is different: it permits a full spectrum of probabilistic distortion guarantees and yields a Lebesgue---Stieltjes average of the single-letter mutual-information leakage under general mechanisms. Recent machine-learning work has also used information-theoretic bounds to analyze learned representations and reconstruction attacks \cite{ZhaoCTG2020,GuoSS2023}; those results are important for representation privacy, but their utility notions are task accuracy or attack advantage rather than the distortion-distribution constraints studied here.

	In \cite{KKSAllerton2016}, we also allow the mechanisms to be either memoryless (also referred to as \textit{local privacy}) or general. This approach has also been considered in the context of differential privacy (DP) (see for example \cite{KoushaISIT2016,Warner,KLNRS11,Kairouz, Duchi}). In the information theoretic context, it is useful to understand how memoryless mechanisms behave for more general distortion constraints as considered here. Furthermore, even less is known about how general mechanisms behave and that is what this paper aims to do.

	In this paper, we first setup the problem formulation in Section \ref{sec: prelim}. Then, in Section \ref{sec: main results} we present our main results for the asymptotic leakage for general and memoryless mechanisms, under the average distortion cost and complementary CDF bounds on distortion. Finally, we provide all the proofs in Sections \ref{sec: proofs}.
	
	\subsection{Notation}
	Throughout this paper we use $D$ as the distortion value, and $d(\cdot,\cdot)$ to indicate the distortion function used for measuring utility. We also use $D_{\text{KL}}(\cdot||\cdot)$ for the KL-divergence between two distributions. The mutual information between two variables $X$ and $Y$ is denoted by $I(X;Y)$ and the base for all the logarithm and exponential functions are the same, but can be any numerical value. We denote binary entropy by $H_b(\cdot)$, and use $\mathbb{E}_{P}[\cdot]$ for expectation with respect to distribution $P$, where the subscript $P$ is dropped when it is clear from context. We denote random variables with capital letters, and their corresponding alphabet set by calligraphic letters. The lower convex envelope of a function $r(\cdot)$ for any point $t$ in its domain is given by
	
	\begin{equation}
	r^{**}(t) \triangleq  \sup\left\{  s(t) { \bigg | } 
	\begin{array}{ll}
	&{ s \text{ is convex}},\\
	&\normalsize{ s(x) \le r(x), \forall x \in \text{Dom } r} 
	\end{array}
	\right\}.
	\end{equation}

	\section{Problem Definition and Preliminaries \label{sec: prelim}}
	Let the source data $(X^n,Y^n)$ be a dataset of $n$ independently and identically distributed (i.i.d.) random variables, where $(X_i,Y_i) \sim P_{X,Y}$, for all $i=1,\ldots,n$. The revealed data is an $n$-length sequence $\hX^n$ drawn from the alphabet $\hat{\mathcal{X}}^n$, and all the alphabet sets $\mc{X}, \mc{Y},\mc{\hX}$ are assumed to be finite sets. A random mechanism is used to generate the revealed data $\hX^n$ given the source data $(X^n,Y^n)$.
	
	In order to quantify the utility of the revealed data, consider the single letter distortion measure as a function $d: \mathcal{X} \times \hat{\mathcal{X}} \rightarrow [D_{\text{min}},D_{\text{max}}]$. Then, the distortion between $n$-length sequences is given by $d(x^n,\hx^n) = \frac{1}{n}\sum_{i=1}^{n} d(x_i,\hx_i)$. 
	The following definitions represent our main quantities of interest, given by the minimum leakage for a dataset subject to a distortion cost constraint and a complementary CDF bound on distortion.
	We differentiate between the memoryless and general mechanisms by the superscripts $M$ and $G$, respectively.
	
	\begin{definition}[Information Leakage under a Cost Function]
		Given a left-continuous and non-decreasing cost function $f: [D_{\text{min}},D_{\text{max}}] \rightarrow [0, \infty)$ and $t > f(D_\text{min})$, the minimal leakage under an expected distortion cost constraint is defined as follows:
		%
		\begin{equation}
		L^{(\cdot)}(n,t,f) \triangleq \min_{\substack{P_{\hX^n|X^n,Y^n} :\\ \mathbb{E}[f(d(X^n,\hat{X}^n))]\le t}} \frac{1}{n}I(Y^n;\hat{X}^n), \label{eq: LG definition}
		\end{equation}
		and
		\begin{equation}
		L^{(\cdot)}(t,f) \triangleq \lim_{n \rightarrow \infty} L^{(\cdot)}(n,t,f),
		\end{equation}
		when the limits exist. The superscript $(\cdot)$ takes values $M$ or $G$, where for $L^{(M)}$ the $n$-letter mechanism $P_{\hX^n|X^n,Y^n}$ is restricted to be stationary and memoryless, i.e. given by $P_{\hX^n|X^n,Y^n} = (P_{\hX|X,Y})^n$, but for $L^{(G)}$ it can be any mechanism satisfying the distortion constraint.
	\end{definition}
	
	
	\begin{definition}[Information Leakage with Distortion CDF Bound]
		Given a right-continuous and non-increasing function $g: [D_{\text{min}}, D_{\text{max}}] \rightarrow (0,1] $, the minimal leakage with a cumulative distortion distribution bounded by $g$ is defined as follows:
		\begin{equation}
		L^{(\cdot)}(n,g) \triangleq \min_{\substack{P_{\hX^n|X^n,Y^n} : \\ \mathbb{P}[d(X^n,\hX^n) > D] \le g(D), \forall D }} \frac{1}{n}I(Y^n;\hat{X}^n),
		\label{eq: g bounded nonasymptotic leakage}
		\end{equation}
		and
		\begin{equation}
		L^{(\cdot)}(g) \triangleq \lim_{n \rightarrow \infty} L^{(\cdot)}(n,g),
		\label{eq: g bounded asymptotic leakage}
		\end{equation}
		when the limits exist. The superscript $(\cdot)$ takes values $M$ or $G$, where for $L^{(M)}$ the $n$-letter mechanism $P_{\hX^n|X^n,Y^n}$ is restricted to be stationary and memoryless, i.e. given by $P_{\hX^n|X^n,Y^n} = (P_{\hX|X,Y})^n$, while for $L^{(G)}$ it can be any mechanism satisfying the distortion constraint.
		\label{def: g bounded asymptotic leakage}
	\end{definition}
	
	We now define the optimal single letter information leakage under a constraint on the expected value of the distortion. This is analogous to the single-letter rate-distortion function, and has appeared in earlier works on privacy \cite{Sankar_TIFS_2013}. As we will show later, this quantity appears as a key element in first-order leakage.
	\begin{definition}[Single Letter Information Leakage]
		\begin{align}
		L(D) &\triangleq \min_{P_{\hat{X}|X,Y} : \mathbb{E}\left[ d(X,\hat{X})\right] \le D} I(Y;\hat{X})\label{eq: L}.
		\end{align}
	\end{definition}
	Note that $L(\cdot)$ is convex, non-increasing, and thus continuous on $(D_{\text{min}},D_{\text{max}}]$. 
	\begin{remark}
		For $f(D)=D$, and any $n$, the optimization in \eqref{eq: LG definition} reduces to \eqref{eq: L} for both memoryless and general mechanisms. 
	\end{remark}
	
	We now define functions that will be critical in expressing asymptotic leakage with the expected distortion cost bound under stationary memoryless and general mechanisms.
	
	\begin{definition}
		For any cost function $f$, and a distortion cost threshold $t > f(D_{\text{min}})$, let
		\begin{align}
		f^{-1}_l(t) \triangleq \sup \{D \in [D_{\text{min}}, D_{\text{max}}]: f(D) < t\},\label{eq: f inv l}\\
		f^{-1}_u(t) \triangleq \sup \{D \in [D_{\text{min}}, D_{\text{max}}]: f(D) \le t\}, \label{eq: f inv h}
		\end{align}
		and define
		\begin{equation}
		\mc{T}_{f} \triangleq \{t: f^{-1}_l(t) \neq f^{-1}_u(t)  \}.
		\label{eq: T_f definition}
		\end{equation}
		Consequently, for any $t \notin \mc{T}_{f}$, we have $f^{-1}_l(t) = f^{-1}_u(t)$, and thus, the inverse function for $f$ can be uniquely determined as
		\begin{equation}
		f^{-1}(t) \triangleq f^{-1}_l(t) = f^{-1}_u(t).
		\end{equation}
	\end{definition}

	\section{Main Results \label{sec: main results}}
	\subsection{Distortion Cost Constraint}
	\begin{theorem}
		Let $t > f(D_{\text{min}})$. If $t \notin \mc{T}_f$, then the asymptotic minimum leakage under stationary memoryless mechanisms is given by
		\begin{equation}
		L^{(M)}(t,f) = (L \circ f^{-1})(t),
		\label{eq: first order leakage memoryless 1}
		\end{equation}
		and for any $t \in \mc{T}_f$, we have
		
		
			\begin{equation}
			L\left(f^{-1}_u(t)\right) \le L^{(M)}(t,f) \le L\left(f^{-1}_l(t)\right),
			\label{eq: first order leakage memoryless 2}
			\end{equation}
			\label{theorem: f approximation memoryless}
		\end{theorem}
	
		\textit{Proof sketch:}
		From the law of large numbers, applying a memoryless mechanism concentrates the distortion around a particular $D$, typically around its expected value, as $n \rightarrow \infty$. Therefore, the distortion cost constraint roughly translates to choosing an expected distortion $D$ such that $f(D) \le t$, or equivalently $f^{-1}(t) \ge D$. If the generalized inverse is unique, then we obtain the exact value in the form of $L(f^{-1}(t))$. At the countable set $\mc{T}_f$ where the inverse jumps, the same argument yields the single-letter upper and lower bounds in \eqref{eq: first order leakage memoryless 2}. For a more detailed proof, see Section \ref{proof: f approximation memoryless}.

		\begin{remark}
		If $f(\cdot)$ is strictly increasing, then $\mc{T}_f = \emptyset$, and $L^{(M)}(t,f)$ is given by \eqref{eq: first order leakage memoryless 1} for any $t$. 
	\end{remark}
	
	\begin{remark}
		For any $t > f(D_\text{min})$, since the closure of the convex hull of epigraphs of $L \circ f^{-1}_l$ and $L \circ f^{-1}_u$ are equal, their lower convex envelopes are equal too. Therefore, $(L \circ f^{-1}_l)^{**} (t) = (L \circ f^{-1}_u)^{**} (t)$, and we refer to this value as $(L \circ f^{-1})^{**} (t)$.
		\label{remark: upper and lower g** are equal}
	\end{remark}
	
	\begin{theorem}
		For $t > f(D_{\text{min}})$, the asymptotic minimum leakage under general mechanisms is given by
		\begin{equation}
		L^{(G)}(t,f) = (L \circ f^{-1})^{**} (t).
		\label{eq: first order leakage}
		\end{equation}
		\label{theorem: f approximation}
	\end{theorem}
	\textit{Proof sketch:} Since $L^{(G)}(t,f)$ is convex in $t$, a convex combination of any two feasible mechanisms is also feasible. Hence, we can always design convex combinations of memoryless mechanisms to achieve the lower convex envelope of $(L \circ f^{-1}) (t)$, and therefore $L^{(G)}(t,f) \le (L \circ f^{-1})^{**} (t)$. Conversely, we show that it is not possible to achieve a smaller leakage. For proof details, we refer the reader to Section \ref{proof: f approximation}.
	\begin{remark}
		Note that for $t\ge f(D_{\text{max}})$, we have $L^{(M)}(t,f)=L^{(G)} (t,f)=0$, where the minimum is achieved by any mechanism with output independent from the input.
		\label{remark: for large t}
	\end{remark}
	\begin{remark}
		If $f$ is convex, then for $t>f(D_{\text{min}})$ we have $(L \circ f^{-1})^{**} (t) = L(f^{-1}(t))$. Therefore, from Theorem \ref{theorem: f approximation memoryless} we have
		\begin{equation}
		L^{(G)}(t,f) = L^{(M)}(t,f) = L(f^{-1}(t)).
		\end{equation}
	\end{remark}
	
	\begin{remark}
		Note that if $L(f^{-1}(t))$ is not equal to its lower convex envelope for some $t$, then the optimal mechanism is formed by a convex combination of the optimal memoryless mechanisms for distortion costs $t_1$ and $t_2$, where $t_1$ is the largest threshold smaller than $t$ and $t_2$ is the smallest threshold larger than $t$, such that $L(f^{-1}(\cdot))$ is equal to its lower convex envelope at $t_1$ and $t_2$.
	\end{remark}
	
	\subsection{Complementary CDF Bound}
	We now proceed to the result on information leakage with distortion CDF bound. In the following, we give closed form results for the asymptotic information leakage with the distortion CDF bounded by a function $g$.
	
	\begin{theorem}
		For a non-increasing right-continuous function $g: [D_{\text{min}}, D_{\text{max}}] \rightarrow (0,1] $, the asymptotic information leakage for memoryless mechanisms under distortion CDF bounded by $g(\cdot)$ is given by
		\begin{equation}
		L^{(M)}(g) = L(D_g),
		\end{equation}
		where $D_g \triangleq \inf \{D \in [D_{\text{min}},D_{\text{max}}]: g(D)<1\}$.
	\end{theorem}
	\begin{IEEEproof}
		Suppose $D_g > D_{\text{min}}$. Then, for any fixed $\delta >0$ and $n$, choose $P_{\hX^n|X^n,Y^n} = \left( P^{*^{(n)}}_{\hX|X,Y} \right ) ^{n}$, where $P^{*^{(n)}}_{\hX|X,Y}$ is the optimal single letter mechanism achieving $L(D_g - \delta)$. Note that by definition $g$ is bounded away from zero, because it is right continuous and considered to be positive over $[D_{\text{min}},D_{\text{max}}]$. Therefore, $\mathbb{P}[d(X^n, \hX^n) > D_g]$ goes to zero as $n$ goes to infinity and the distortion constraint $\mathbb{P}[d(X^n, \hX^n) > D] \le g(D)$ is satisfied for all $D$ for sufficiently large $n$. Then, as $\delta \rightarrow 0$, continuity of $L(\cdot)$ implies $L(D_g)$ is achievable.
		
		Conversely, according to the law of large numbers, the distortion $d(X^n, \hX^n)$ concentrates around its expected value as $n$ goes to infinity. In other words, we have $\mathbb{P}[d(X^n,\hX^n) > D] \rightarrow 1$, if $D < \mathbb{E}[d(X^n, \hX^n)]$. This, in turn, implies that for any $D$ such that $g(D)<1$, we must have $\mathbb{E} [d(X^n, \hX^n)] \le D$. Therefore, a feasible memoryless mechanism has to satisfy $\mathbb{E}[d(X^n, \hX^n)] \le D_g$.
		
		Finally, for $D_g=D_{\text{min}}$, we have to satisfy $\mathbb{P}[d(X^n, \hX^n) = D_{\text{min}}]=1$. Note that in this case, the constraint for $L(D_g)$, i.e. $\mathbb{E}[d(X^n, \hX^n)] \le D_g$, is also equivalent to $\mathbb{P}[d(X^n, \hX^n) = D_{\text{min}}]=1$. Therefore, the set of feasible memoryless mechanisms for $L^{(M)}(g)$ is equal to those for $L(D_g)$, and thus, $L^{(M)}(g)=L(D_g)$.
	\end{IEEEproof}

	\begin{theorem}
		Let $g: [D_{\text{min}}, D_{\text{max}}] \rightarrow (0,1] $ be a non-increasing right-continuous function. If the single letter leakage function $L(\cdot)$ is bounded on $[D_{\text{min}}, D_{\text{max}}]$, then the asymptotic information leakage for general mechanisms under distortion CDF bounded by $g(\cdot)$ is given by
		\begin{equation}
		L^{(G)}(g)= \int_{D_{\text{min}}}^{D_{\text{max}}} L(D) d(g(D)),
		\label{eq: g bounded asymptotic leakage result}
		\end{equation}
		where the integral is a Lebesgue–--Stieltjes integral of the single letter leakage function $L(\cdot)$ with respect to the Lebesgue–--Stieltjes measure associated with the constraint function $g$.
		\label{theorem: g bounded asymptotic leakage}
	\end{theorem}
	
	\textit{Proof sketch:}
	We first prove this result for simple constraint functions $g$, which are in the form of a finite sum of step functions. Then, we show that any non-increasing right-continuous constraint function $g$ can be upper and lower bounded by such simple functions, and therefore, the corresponding leakage can be upper and lower bounded by that of the simple functions. For a more detailed proof, see Section \ref{proof: g bounded asymptotic leakage}.
	
	\begin{remark}
		An alternative way of describing the result in Theorem \ref{theorem: g bounded asymptotic leakage} is that the asymptotically optimal mechanism behaves as if it first chooses a random $D$ drawn from a distribution with a complementary CDF exactly equal to $g(\cdot)$, and then applies the single letter optimal mechanism achieving the single letter optimal leakage $L(D)$ in a stationary and memoryless fashion.
		Thus, averaging over the random choice of $D$, the resulting leakage is given as the integral in \eqref{eq: g bounded asymptotic leakage result}.
	\end{remark}
	
	\subsection{Auxiliary Result}
	We now present a result characterizing the asymptotic optimal privacy leakage subject to multiple excess probability constraints. This can be seen as a special case of complementary CDF bound in which the $g$ function is a simple function, i.e. it takes finitely many values. The following results will also be used in the proof of Theorem \ref{theorem: f approximation}.
	
\begin{figure}[htb!]
		\centering
		\includegraphics[width= 0.4 \columnwidth]{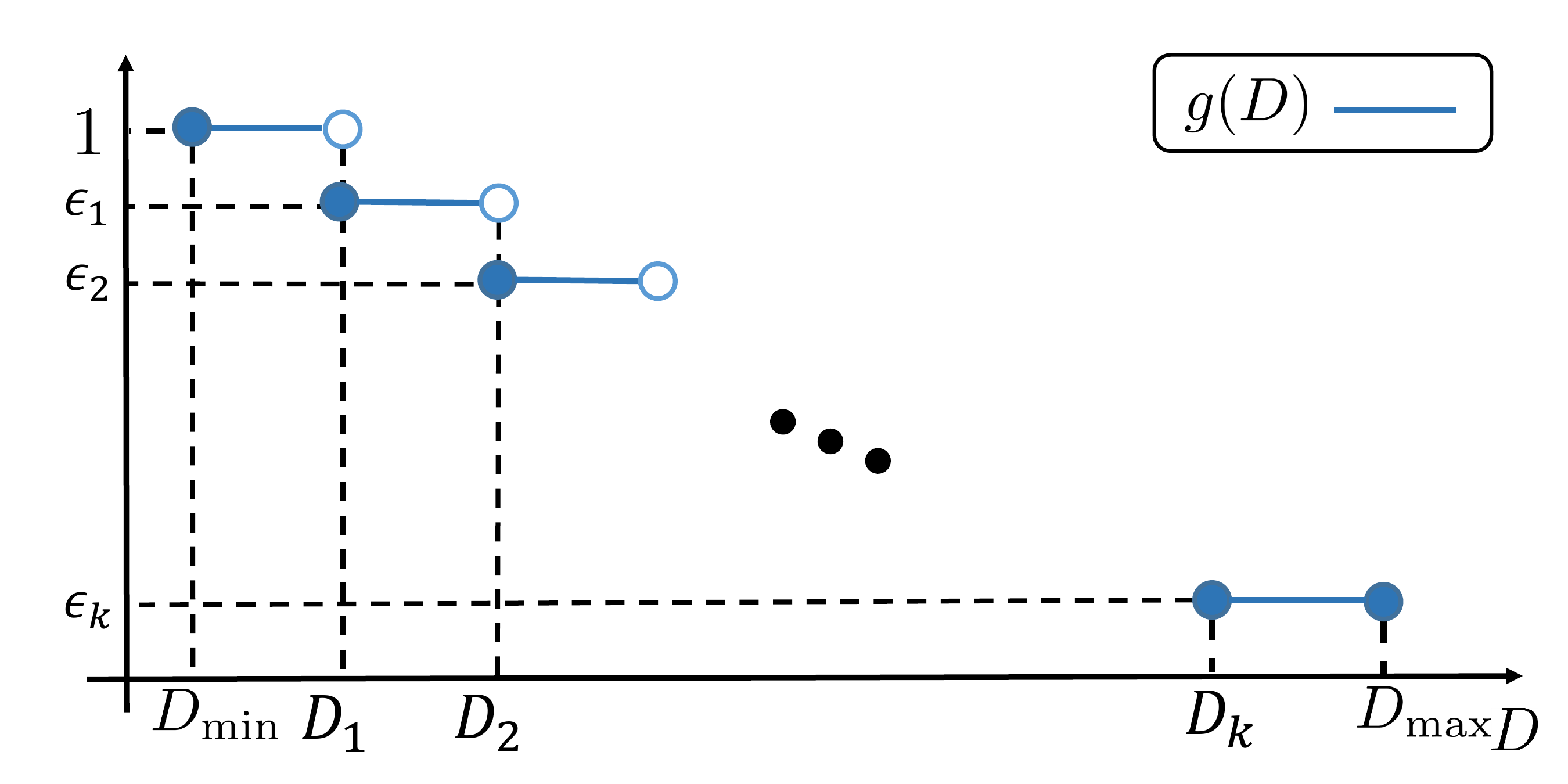}
		\caption{A simple $g_{\boldsymbol{\epsilon},\boldsymbol{D}}(D)$.}
		\label{fig: simple g function}
		\vspace{-10pt}
	\end{figure}

	For vectors $\boldsymbol{D} = (D_1, D_2, \ldots, D_k)$ and $\boldsymbol{\epsilon} = (\epsilon_1, \epsilon_2, \ldots, \epsilon_k)$, where $D_{\text{min}} \le D_1 < \cdots < D_k \le D_{\text{max}}$ and $1 \ge \epsilon_1 > \cdots > \epsilon_k > 0$, a simple function $g_{\boldsymbol{\epsilon},\boldsymbol{D}}$ is illustrated in Fig. \ref{fig: simple g function} and formally defined as
	\begin{equation}
	g_{\boldsymbol{\epsilon},\boldsymbol{D}}(D) \triangleq \begin{cases}
	1, & D_{\text{min}} \le D < D_1,\\
	\epsilon_i, & D_{i} \le D < D_{i+1}, i=1,\ldots,k-1,\\
	\epsilon_k, & D_k \le D \le D_{\text{max}}.
	\end{cases}
	\label{eq: simple function}
	\end{equation}
	One can verify that for a constraint function of this form, the minimization in \eqref{eq: g bounded asymptotic leakage} is equivalent to the \textit{information leakage with multiple excess distortion constraints}, defined as follows.
	\begin{definition}[Information Leakage with Multiple Excess Probability Constraints]
		Given a distortion vector $\boldsymbol{D} = (D_1, D_2, \ldots, D_k)$ and a tail probability vector $\boldsymbol{\epsilon} = (\epsilon_1, \epsilon_2, \ldots, \epsilon_k)$, where $D_{\text{min}} \le D_1 < \cdots < D_k \le D_{\text{max}}$ and $1 \ge \epsilon_1 > \cdots > \epsilon_k > 0$, the minimal leakage with multiple excess distortion constraints is defined as
		\begin{align}
		L^{(G)}(n,\boldsymbol{D},\boldsymbol{\epsilon}) &\triangleq \min_{\substack{P_{\hX^n|X^n,Y^n} :\\ \mathbb{P}[d(X^n,\hX^n) > D_i] \le \epsilon_i, \\ \forall 1 \le i  \le k}} \frac{1}{n}I(Y^n;\hat{X}^n), 
		\label{eq: multi constraint}
		\end{align}
		where the $n$-letter mechanisms in \eqref{eq: LG definition} are not constrained to be memoryless or stationary, and
		\begin{equation}
		L^{(G)}(\boldsymbol{D},\boldsymbol{\epsilon}) \triangleq \lim_{n \rightarrow \infty} L^{(G)}(n,\boldsymbol{D},\boldsymbol{\epsilon}),
		\end{equation}
		\label{def: multi constraint}
	\end{definition}
	when the limit exists. In the following lemma, we provide the asymptotic optimal leakage under general mechanisms for the class of distortion CDF bound functions defined in Definition \ref{def: multi constraint}.
	\begin{lemma}
		\begin{align}
		L^{(G)}(\boldsymbol{D},\boldsymbol{\epsilon})	&= \sum_{i=1}^{k} (\epsilon_{i-1}-\epsilon_i) L(D_i) \nonumber\\
		&= \int_{D_{\text{min}}}^{D_{\text{max}}} L(D) d(g_{\boldsymbol{\epsilon},\boldsymbol{D}}(D)),
		\label{eq: multi constraint result}
		\end{align}
		where $\epsilon_0 = 1$. In particular, we have
		\begin{equation}
		L^{(G)}(n,\boldsymbol{D},\boldsymbol{\epsilon}) = \sum_{i=1}^{k} (\epsilon_{i-1}-\epsilon_i) L(D_i) + \theta (k,n),
		\end{equation}
				where
				\edit{}{
					\begin{subequations}
						\begin{align}
						-\beta_n &\le  \theta (k,n) \le O\left(\sqrt {\frac{\log n}{n}}\right),
						\label{eq: multi constraint LB and UB}\\
						\beta_n &\triangleq \psi\left(\sqrt{\frac{2\log(k+1)}{n}}\right) + \frac{\log(k+1)}{n}.
						\label{eq: beta_n definition}
						\end{align}
					\end{subequations}
				}
				for some non-decreasing concave function $\psi:[0,2]\to[0,\infty)$, depending only on $(P_{X,Y},d)$, such that $\psi(0)=0$. In particular, $\beta_n \to 0$.
			\label{lemma: g bounded asymptotic leakage - simple}
		\end{lemma}
		\begin{remark}
				The remainder term in \eqref{eq: beta_n definition} simplifies further when $\psi$ admits an explicit bound near the origin. If there exist constants $C,\delta_0>0$ and $\alpha\in(0,1]$ such that
				\[
				\psi(u)\le Cu^{\alpha},
				\qquad 0\le u\le \delta_0,
				\]
				then
				\[
				\beta_n = O\left(\left(\frac{\log(k+1)}{n}\right)^{\alpha/2} + \frac{\log(k+1)}{n}\right).
				\]
				In particular, if $\psi$ is locally Lipschitz near the origin, then
				\[
				\beta_n = O\left(\sqrt{\frac{\log(k+1)}{n}}\right).
				\]
		\end{remark}
		
		\textit{Proof sketch:}
		The proof hinges on choosing a combination of memoryless mechanisms, each of them being the single letter optimal mechanism for a separate $D_i$ applied in a stationary and memoryless fashion. The weights of this combination will be chosen such that all the excess distortion probabilities are met. For a detailed proof see section \ref{proof: g bounded asymptotic leakage - simple}. 
	
	\section{Illustration of Results}
	In this section, we first examine the generic cases of single and double step $f$ and $g$ functions. Then, we consider a doubly symmetric binary source and derive its corresponding single letter leakage function. Finally, we use the single letter leakage function to find the asymptotically optimal leakage under specific examples of the average distortion cost constraint and complementary CDF bound.
	\subsection{Distortion Cost Function}
	\begin{example}
		$f(D)=\boldsymbol{1} (D>D_0)$ as shown in Fig. \ref{fig: f one step}. In this case, $\mathcal{T}_f = \{1\}$, and we have
		\begin{figure}[htb!]
			\centering
			\includegraphics[width= 0.4 \columnwidth]{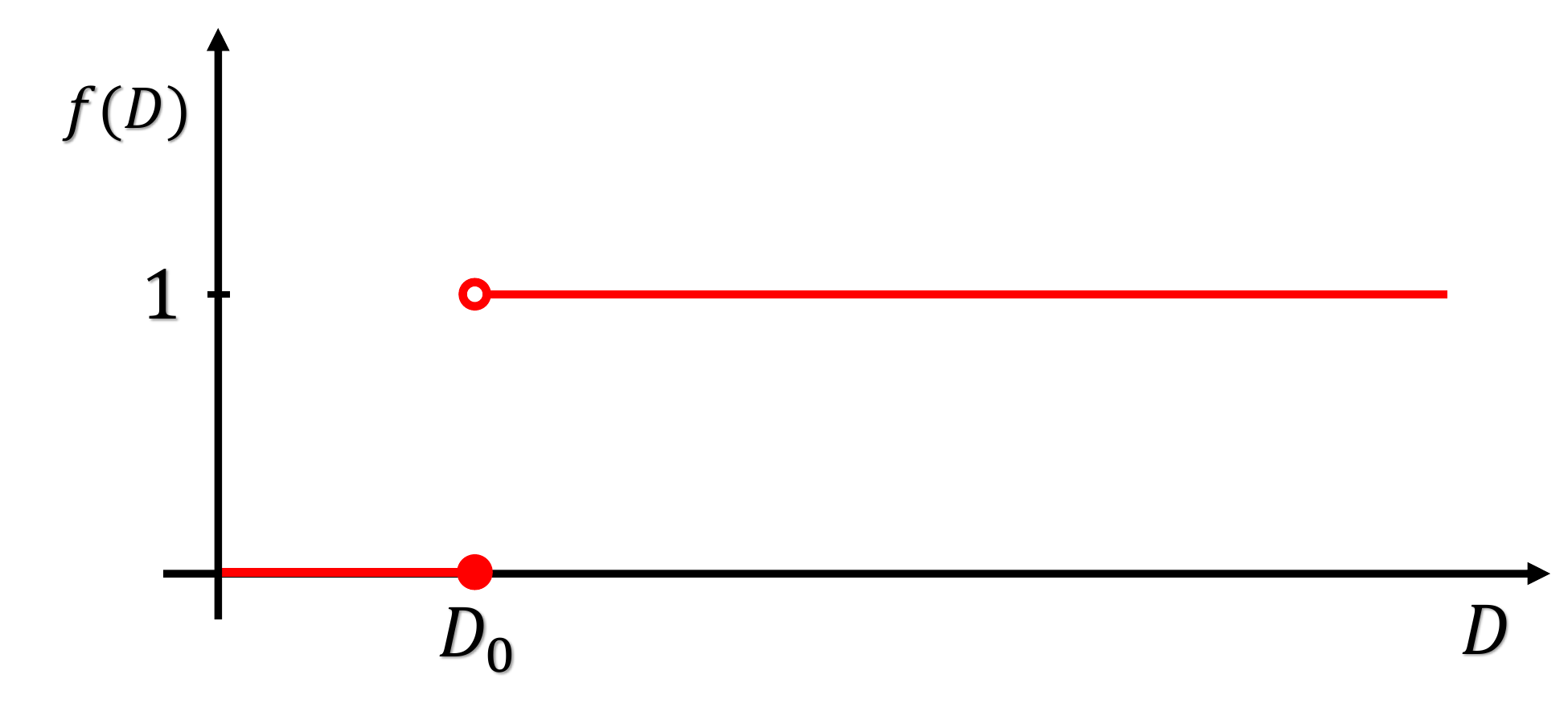}
			\caption{The single step cost function $f(D)=\boldsymbol{1} (D>D_0)$.}
			\label{fig: f one step}
		\end{figure}
		\begin{align}
		f^{-1}_u(t) &= \begin{cases}
		D_0, & t<1,\\
		D_{\text{max}},& t \ge 1,
		\end{cases}\\
		f^{-1}_l(t) &= \begin{cases}
		D_0, & t \le 1,\\
		D_{\text{max}},& t > 1.
		\end{cases}
		\end{align}
		Therefore, according to Theorem \ref{theorem: f approximation memoryless} for stationary memoryless mechanisms we have
		\begin{equation}
		L^{(M)}(t,f) = \begin{cases}
		L({D_0}), &  0 < t < 1,\\
		0,& t\ge 1,
		\end{cases}
		\end{equation}
		and for general mechanisms, according to Theorem \ref{theorem: f approximation} we have
		\begin{equation}
		L^{(G)}(t,f) = \begin{cases}
		(1-t) L({D_0}), & 0 \le t < 1\\
		0, & t \ge 1.
		\end{cases}
		\end{equation}
		This exactly matches our earlier results in \cite{KKSAllerton2016} and for the special case of $X=Y$ simplifies to the result in \cite{Kostina}. The leakages $L^{(G)}$ and $L^{(M)}$ are depicted in Fig. \ref{fig: example2}.
		Note that for $t=1$, we have $L^{(G)}(t,f) = L^{(M)}(t,f) = 0$ due to Remark \ref{remark: for large t}.
		\begin{figure}
			\centering
			\includegraphics[width= 0.4 \columnwidth]{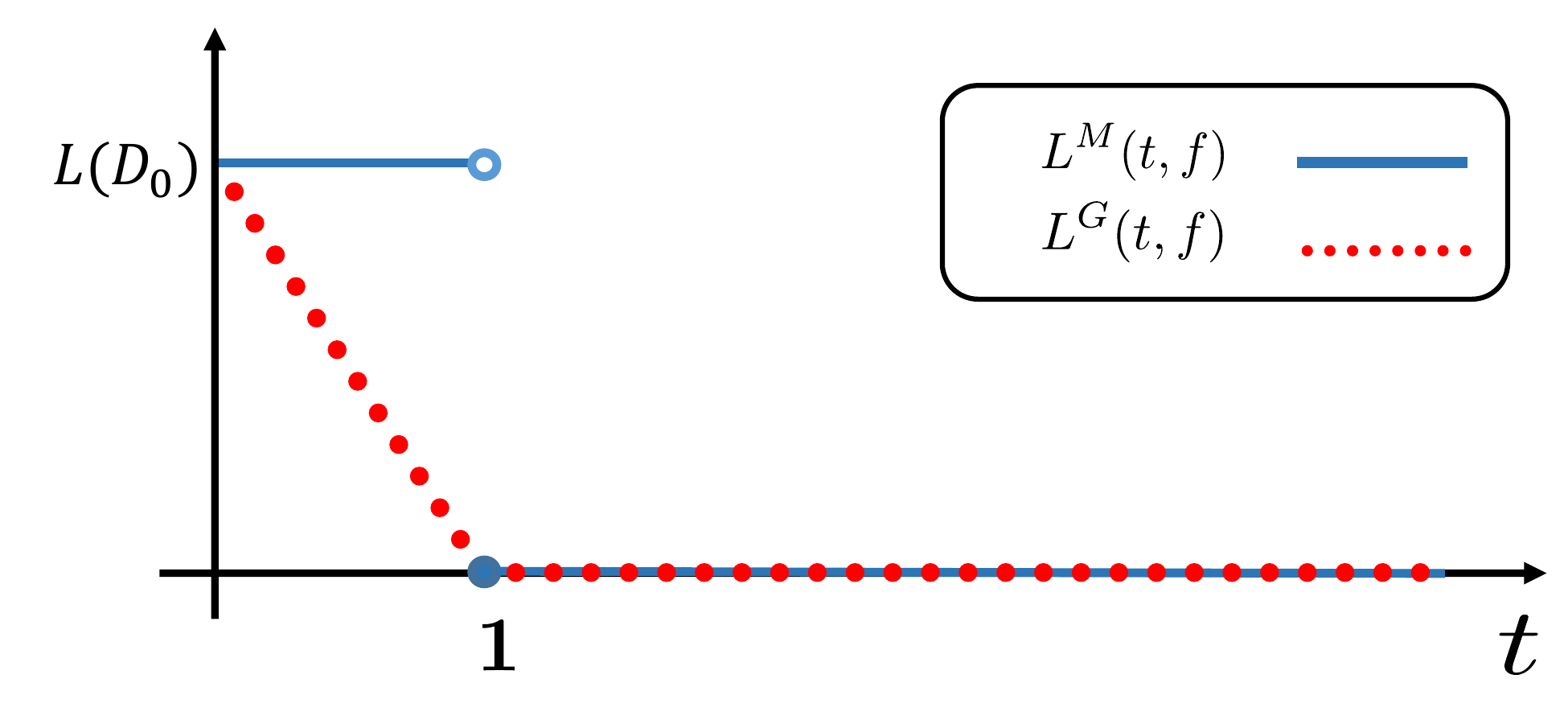}
			\caption{The leakage functions $L^{(M)}(t,f)$ and $L^{(G)}(t,f)$ for $f(D)=\boldsymbol{1} (D > D_0)$.}
			\label{fig: example2}
			\vspace{-10pt}
		\end{figure}
		\label{example: 1}
	\end{example}
	\begin{example}
		$f(D)=a_1 \boldsymbol{1} (D>D_1) + a_2 \boldsymbol{1} (D > D_2)$, $D_1 < D_2$ as shown in Fig. \ref{fig: f two step}. In this case, $\mathcal{T}_f = \{a_1,a_1+a_2\}$, and we have
		\begin{align}
		f^{-1}_u(t) &= \begin{cases}
		D_1, & t<a_1,\\
		D_2, & a_1 \le t < a_1+ a_2,\\
		D_{\text{max}},& t \ge a_1 + a_2,
		\end{cases}\\
		f^{-1}_l(t) &= \begin{cases}
		D_1, & t\le a_1,\\
		D_2, & a_1 < t \le a_1+ a_2,\\
		D_{\text{max}},& t > a_1 + a_2.
		\end{cases}
		\end{align}
		Hence, according to Theorem \ref{theorem: f approximation memoryless} for stationary memoryless mechanisms we have
		\begin{equation}
		L^{(M)}(t,f) = \begin{cases}
		L(D_1), & t < a_1,\\
		L(D_2), & a_1 < t< a_1 + a_2,\\
		0, & a_1 + a_2 \le t.
		\end{cases}
		\end{equation}
			Note that for $t=a_1$, Theorem \ref{theorem: f approximation memoryless} only yields the bounds $L(D_2) \le L^{(M)}(t,f) \le L(D_1)$, while for $t=a_1+a_2$ we have $L^{(M)}(t,f) =0$ due to Remark \ref{remark: for large t}.
		From Theorem \ref{theorem: f approximation}, we know that $L^{(G)}(t,f)$ is the lower convex envelope of $L^{(M)}(t,f)$. If $a_2 L(D_1) \ge  (a_1 + a_2) L(D_2)$, then it is given by
		\begin{equation}
		L^{(G)}(t,f) = \begin{cases}
		L(D_2)+(1 - \frac{t}{a_1}) L(D_1) & t \le a_1,\\
		(1-\frac{t-a_1}{a_2}) L(D_2) & a_1 \le t \le a_1+a_2,\\
		0 & a_1 + a_2 \le t,
		\end{cases}
		\end{equation}
		and otherwise,
		\begin{equation}
		L^{(G)}(t,f) = \begin{cases}
		(1-\frac{t}{a_1+a_2}) L(D_1)& t \le a_1 + a_2 , \\
		0 & a_1 + a_2 \le t.
		\end{cases}
		\end{equation}
		These two cases together with their corresponding $L^{(M)}(t,f)$ are shown in Figs. \ref{fig: example3a} and \ref{fig: example3b}, respectively.
		\begin{figure}[htb!]
			\centering
			\includegraphics[width= 0.4 \columnwidth]{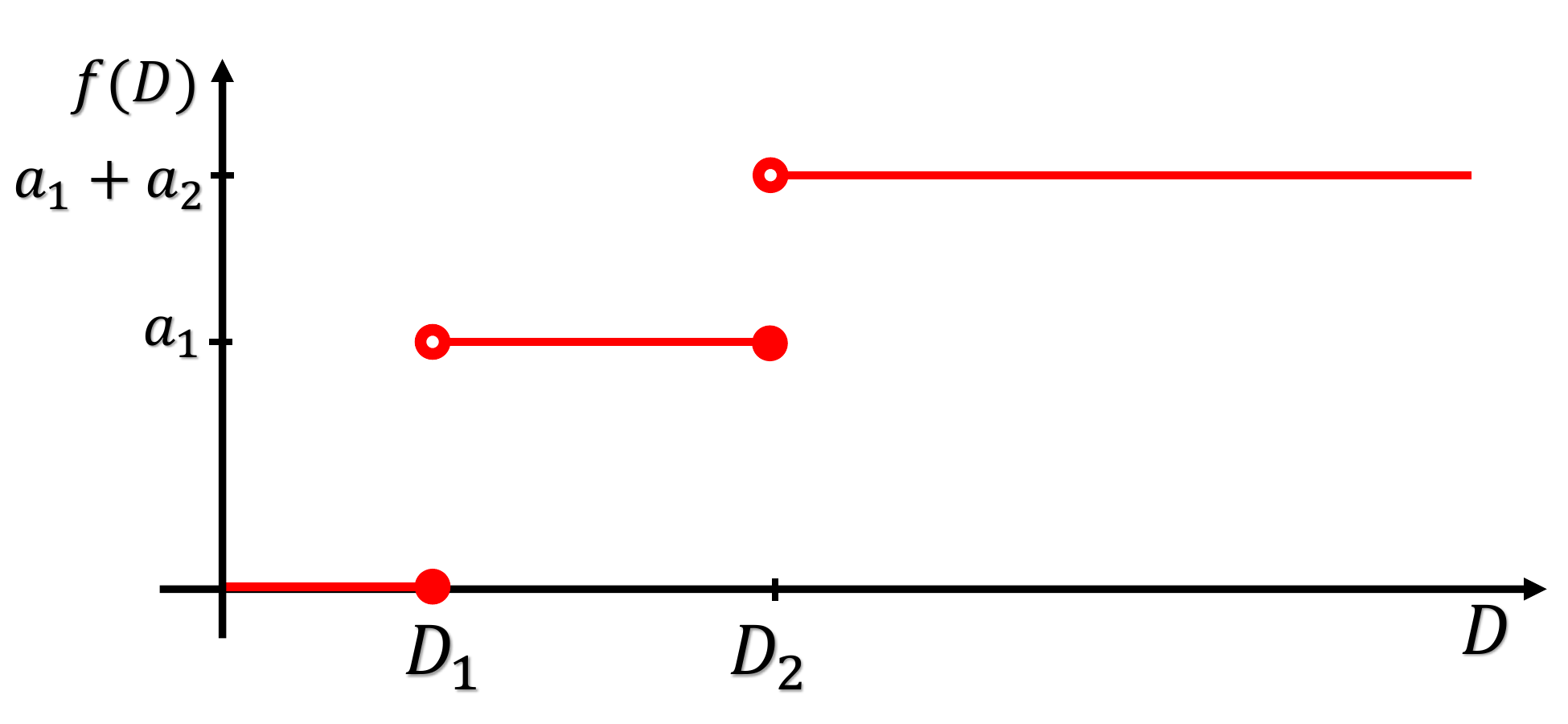}
			\caption{The double step cost function $f(D)=a_1 \boldsymbol{1} (D>D_1) + a_2 \boldsymbol{1} (D > D_2)$, $D_1 < D_2$.}
			\label{fig: f two step}
		\end{figure}
		\begin{figure}[htb!]
			\centering
			\includegraphics[width= 0.4 \columnwidth]{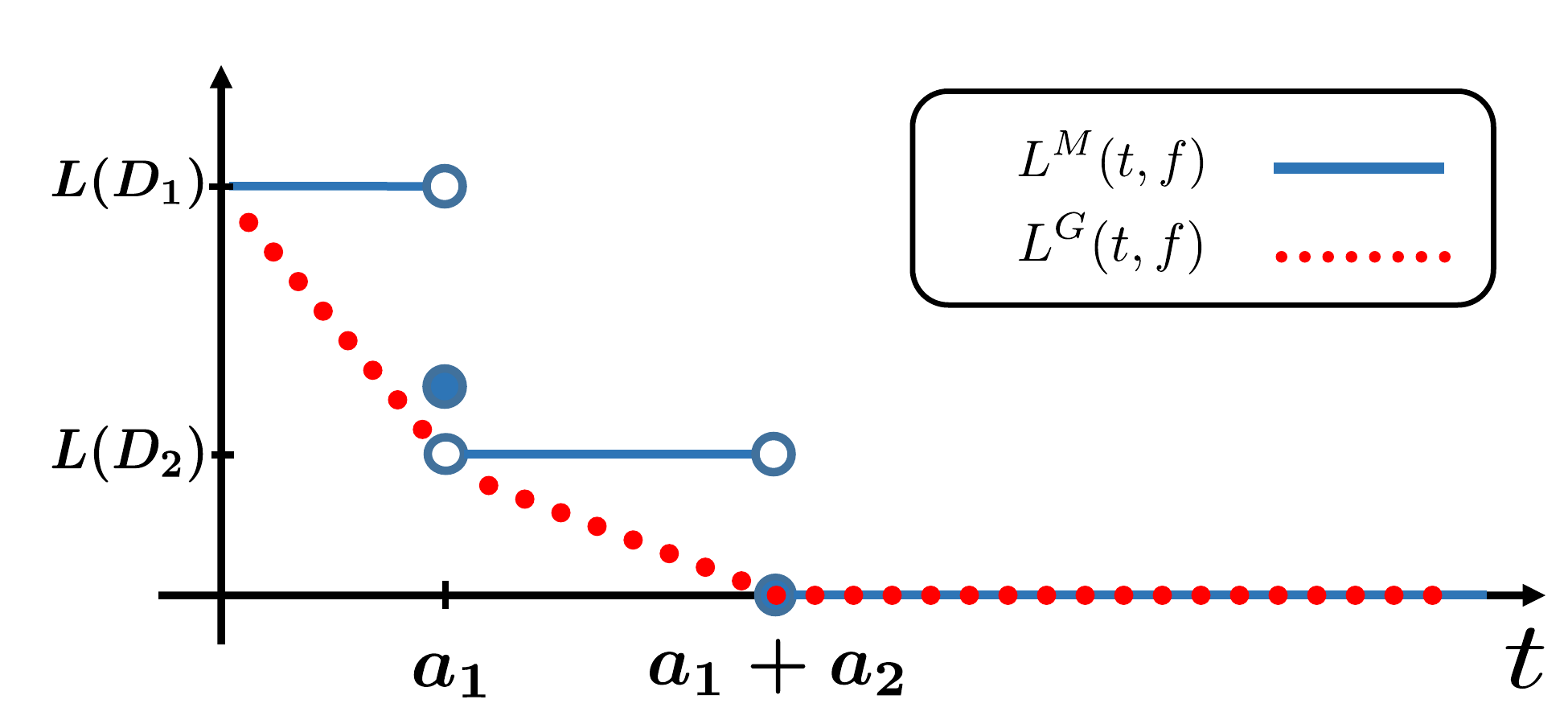}
			\caption{$L^{(M)}(t,f)$ and $L^{(G)}(t,f)$ for $f(D)=a_1 \boldsymbol{1} (D>D_1) + a_2 \boldsymbol{1} (D>D_2)$, if $a_2 L(D_1) \ge  (a_1 + a_2) L(D_2)$.}
			\label{fig: example3a}
		\end{figure}
		\begin{figure}[htb!]
			\centering
			\includegraphics[width= 0.4 \columnwidth]{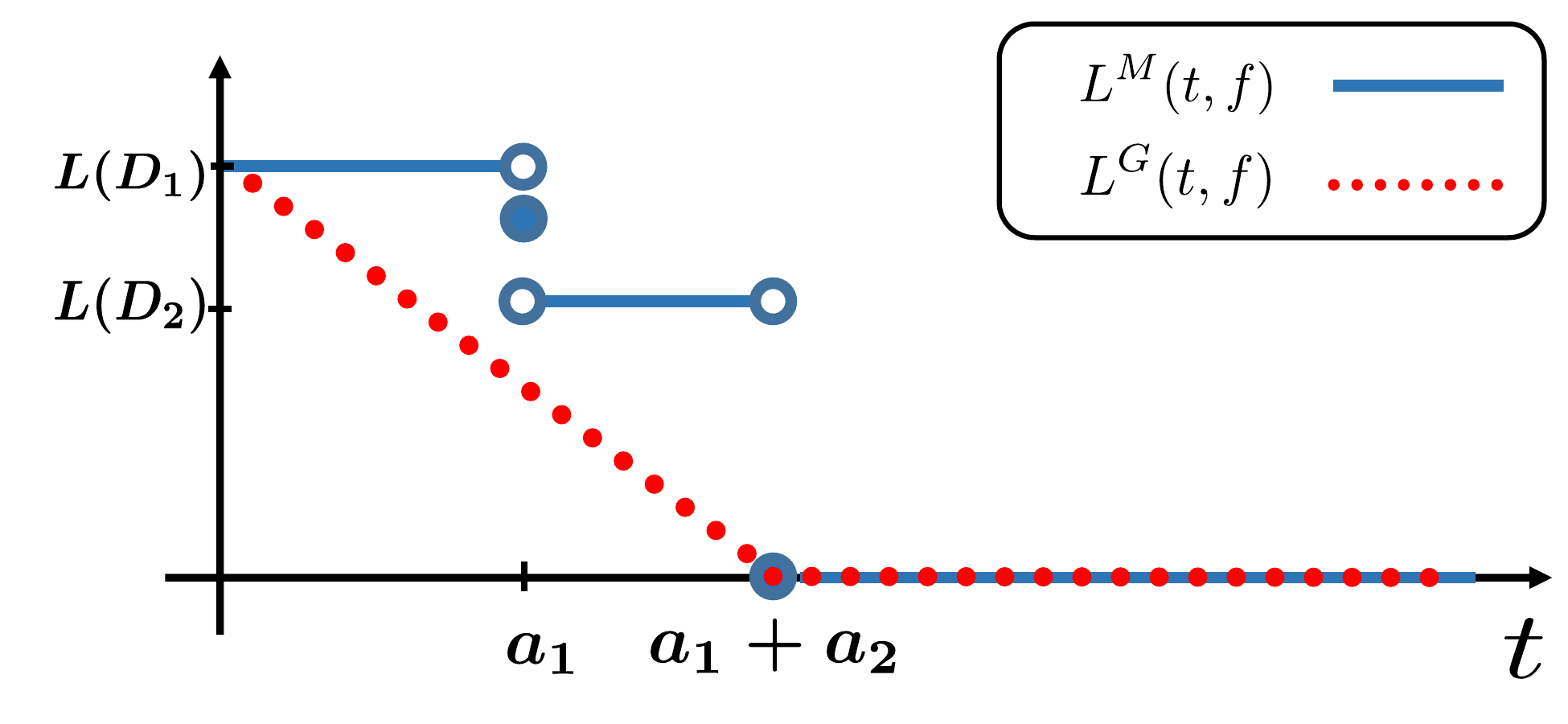}
			\caption{$L^{(M)}(t,f)$ and $L^{(G)}(t,f)$ for $f(D)=a_1 \boldsymbol{1} (D>D_1) + a_2 \boldsymbol{1} (D>D_2)$, if $a_2 L(D_1) <  (a_1 + a_2) L(D_2)$.}
			\label{fig: example3b}
		\end{figure}
	\end{example}
	
	\subsection{Distortion CDF Constraints}
	We now proceed to complementary CDF bounds on distortion. First, we consider a single step function $g$ (hard tail probability constraint), and then generalize to a sum of two step functions.
	
	\begin{example}
		$g(D) = 1- (1-\epsilon) \boldsymbol{1}(D \ge D_0)$ as shown in Fig. \ref{fig: example4}, where $D_{\text{min}}<D_0<D_{\text{max}}$. For stationary memoryless mechanisms we have
		\begin{equation}
		L^{(M)}(g) = L(D_0),
		\label{eq: LM for tail prob const}
		\end{equation}
		while for the general mechanisms, we have
		\begin{equation}
		L^{(G)}(g) = \int_{D_{\text{min}}}^{D_{\text{max}}} L(D) d(g(D)) = (1-\epsilon)L(D_0).
		\label{eq: LG for tail prob const}
		\end{equation}	
		\begin{figure}[htb!]
			\centering
			\includegraphics[width= 0.45 \columnwidth]{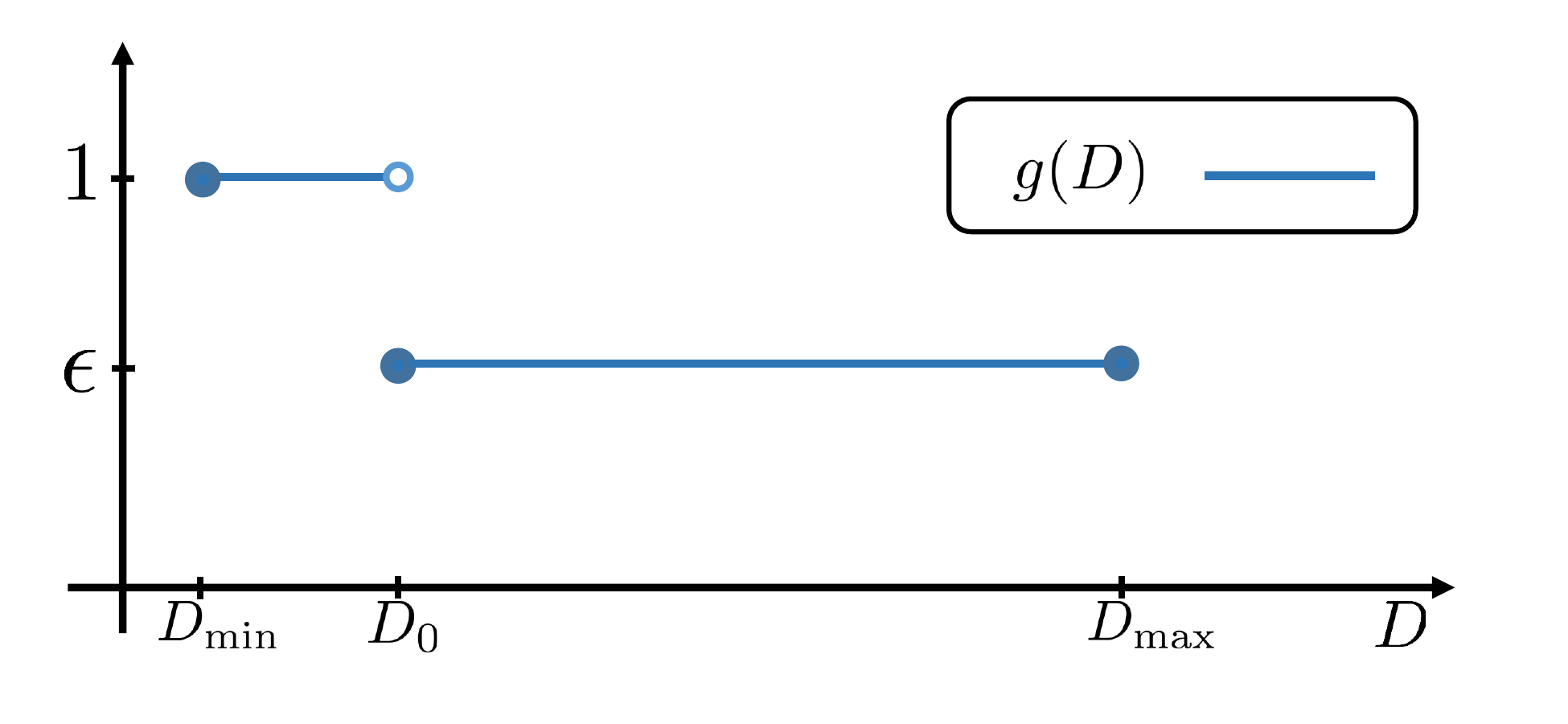}
			\caption{The single step complementary CDF bound function $g(D)=1- (1-\epsilon) \boldsymbol{1}(D \ge D_0)$.}
			\label{fig: example4}
		\end{figure}
		Note that this is equivalent to Example \ref{example: 1}. Therefore, \eqref{eq: LM for tail prob const} and \eqref{eq: LG for tail prob const} verify the results in \cite{Kostina} and \cite{KKSAllerton2016}, wherein the tail probability constraint is used as a utility metric.
	\end{example}
	
	\begin{example}
		$g(D) = \boldsymbol{1}(D < D_1) + \epsilon_1 \boldsymbol{1}( D_1 \le D < D_2) + \epsilon_2 \boldsymbol{1}( D_2 \le D) $ as shown in Fig. \ref{fig: example5}. For stationary memoryless mechanisms we have
		\begin{equation}
		L^{(M)}(g) = L(D_1),
		\label{eq: LM for example 5}
		\end{equation}
		while for the general mechanisms, we have
		\begin{equation}
		L^{(G)}(g) = (1-\epsilon_1)L(D_1) + (\epsilon_1 - \epsilon_2) L(D_2).
		\label{eq: LG for example 5}
		\end{equation}
		\begin{figure}[htb!]
			\centering
			\includegraphics[width= 0.45 \columnwidth]{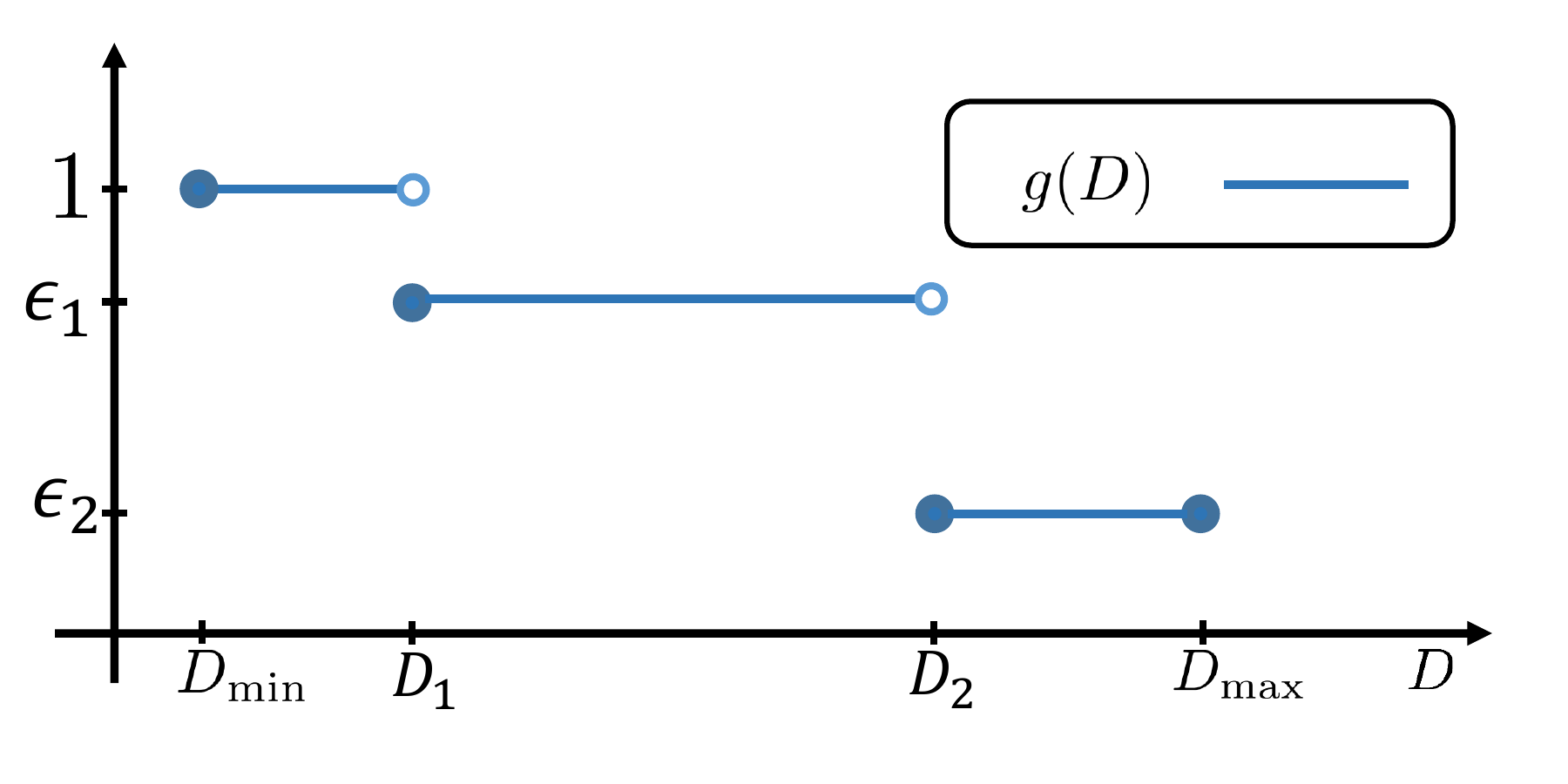}
			\caption{The double step complementary CDF bound function $g(D) = \boldsymbol{1}(D < D_1) + \epsilon_1 \boldsymbol{1}( D_1 \le D < D_2) + \epsilon_2 \boldsymbol{1}( D_2 \le D)$.}
			\label{fig: example5}
		\end{figure}
	\end{example}
	
	\subsection{Doubly Symmetric Binary Source (DSBS)}
\begin{figure}[htb!]
		\centering
		\includegraphics[width= 0.25 \columnwidth]{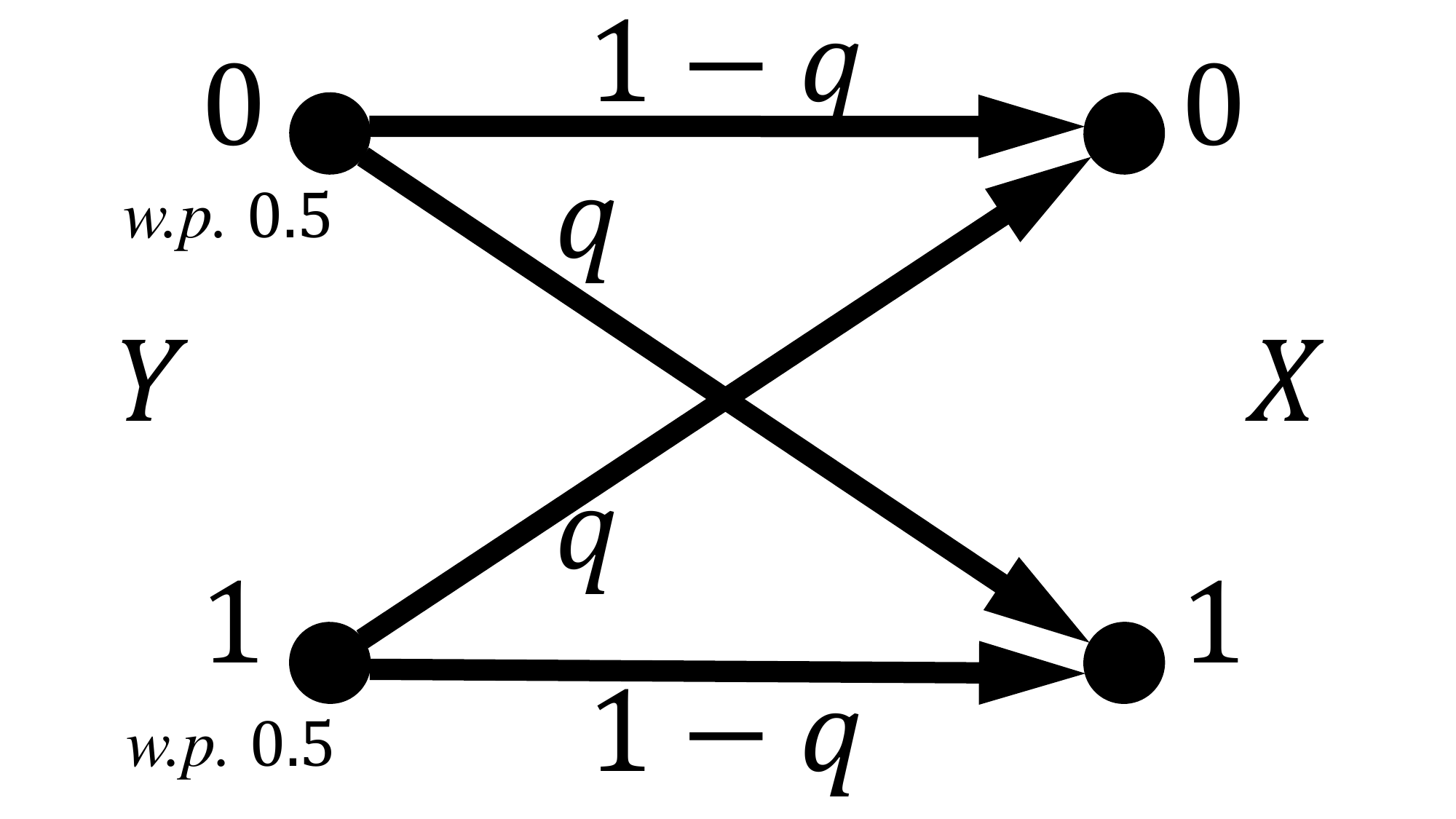}
		\caption{A doubly symmetric source with parameter $q$.}
		\label{fig: binary source}
	\end{figure}

	We now consider a doubly symmetric source with parameter $q$ as depicted in Fig. \ref{fig: binary source} with Hamming distortion, i.e. $d(x,\hx) = \boldsymbol{1}(x \neq \hx)$, as the utility metric. 
	In the following lemma, proved in Section \ref{proof: single letter leakage for doubly symmetric source}, we derive the single letter leakage function for this source.
	\begin{lemma}
		For a doubly symmetric source with $q \le 0.5$, the single letter leakage function is given by
		\begin{equation}
		L(D) = \begin{cases}
		1 - H_b(q + D), & D < 0.5 - q,\\
		0, &D\ge 0.5 - q.
		\end{cases} 
		\label{eq: single letter leakage for doubly symmetric source}
		\end{equation}
		\label{lemma: single letter leakage for doubly symmetric source}
	\end{lemma}
	\begin{remark}
		Due to the inherent symmetry of the problem, for all $q > 0.5$, Lemma \ref{lemma: single letter leakage for doubly symmetric source} holds with $q$ replaced by $1-q$.
	\end{remark}
	
	Given the single letter leakage function for a doubly symmetric source, we provide numerical examples for the asymptotically optimal leakages under both distortion cost constraints and complementary CDF bounds.
	
	\begin{example}
		For a doubly symmetric source with parameter $q=0.1$ and Hamming distortion, consider the cost function
		\begin{equation}
		f(D) = \begin{cases}
		\frac{4(8D - i -0.5)^5+16D+1-2i}{32},& D \in [\frac{i}{8}, \frac{i+1}{8}), i \in \{0,\ldots,7\}, \\
		1,& D=1,
		\end{cases}
		\end{equation}
		as shown in Fig. \ref{fig: f}. Then, the corresponding leakage functions $L^{(M)}(t,f)$ and $L^{(G)}(t,f)$ are shown in Fig. \ref{fig: example f}.
		
		\begin{figure}[htb!]
			\centering
			\includegraphics[width= 0.45 \columnwidth]{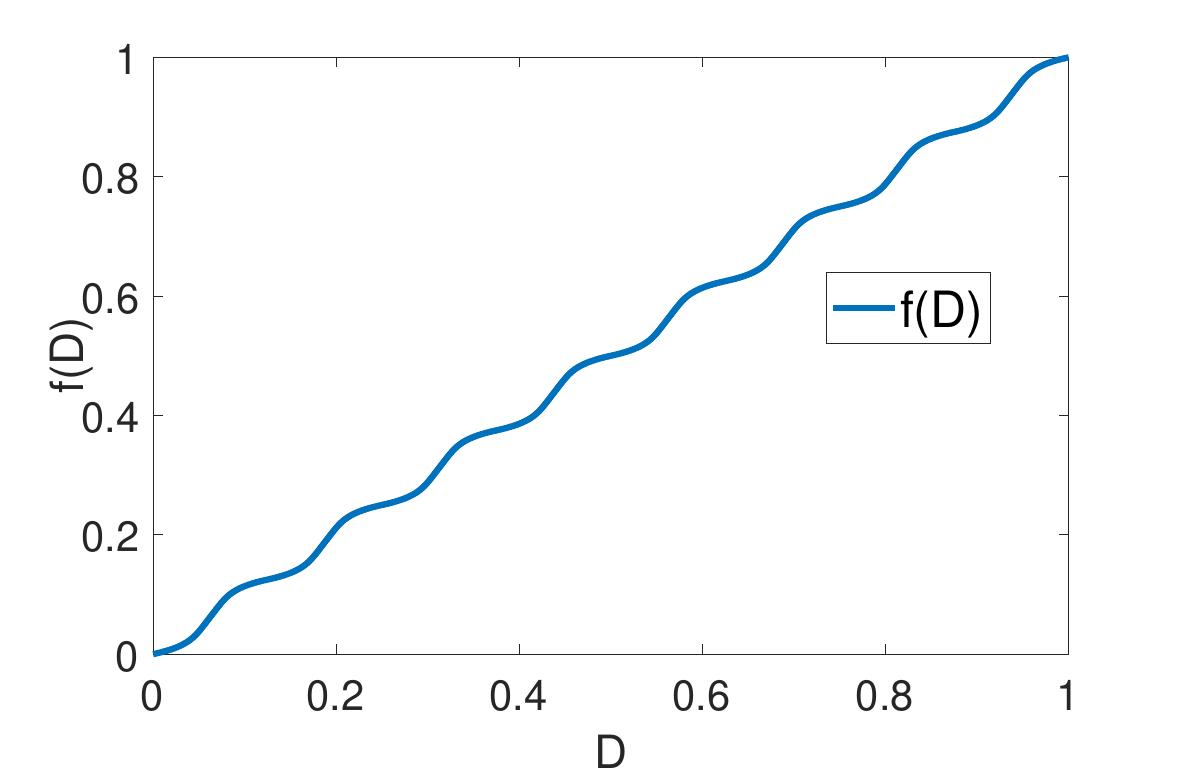}
			\caption{The cost function $f(D)$ for Example \ref{example: 3}.}
			\label{fig: f}
		\end{figure}
		
		\begin{figure}[htb!]
			\centering
			\includegraphics[width= 0.45 \columnwidth]{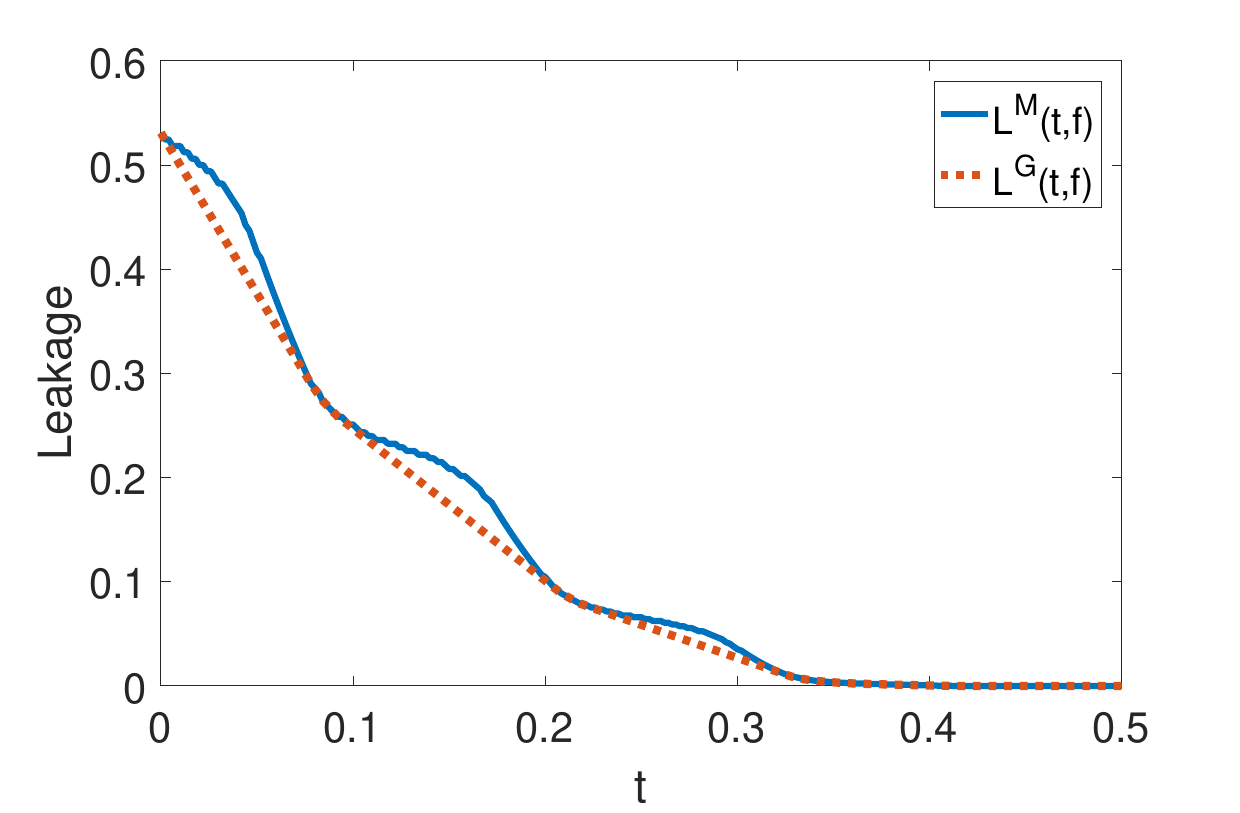}
			\caption{Memoryless and general leakage functions $L^{(M)}(t,f)$ and $L^{(G)}(t,f)$ for Example \ref{example: 3}.}
			\label{fig: example f}
		\end{figure}
		\label{example: 3}
	\end{example}
	
	We now proceed to an examples that resemble a \textit{soft} single step complementary CDF bound. We choose functions that are parametrized with a parameter $\lambda$ such that they converge to a hard single step CDF bound as $\lambda \rightarrow \infty$. 
	
	\begin{example}
		Consider a doubly symmetric source with parameter $q=0.1$. Then, for any $\lambda \ge 0 $ define
		\begin{align}
		g_\lambda(D) =& \epsilon + (1-\epsilon)  \boldsymbol{1} (D \le D_0) \nonumber \\
		& + (1-\epsilon)\left(\frac{1}{2} - \boldsymbol{1} (D \le D_0)\right) e^{-\lambda |D - D_0|}.
		\end{align}
		In Fig. \ref{fig: g soft}, this function is plotted for $D_0 = 0.2$, $\epsilon = 0.1$, and four different values of $\lambda$. Note that in Fig. \ref{fig: LGg}, the value of $L^{(G)}(g_\lambda)$ converges to the asymptotic value of $(1-\epsilon) L(D_0)$ as $\lambda \rightarrow \infty$, and $L^{(G)}(g_\lambda)$ is non-monotonic in $\lambda$.	
		\begin{figure}[htb!]
			\centering
			\includegraphics[width= 0.45 \columnwidth]{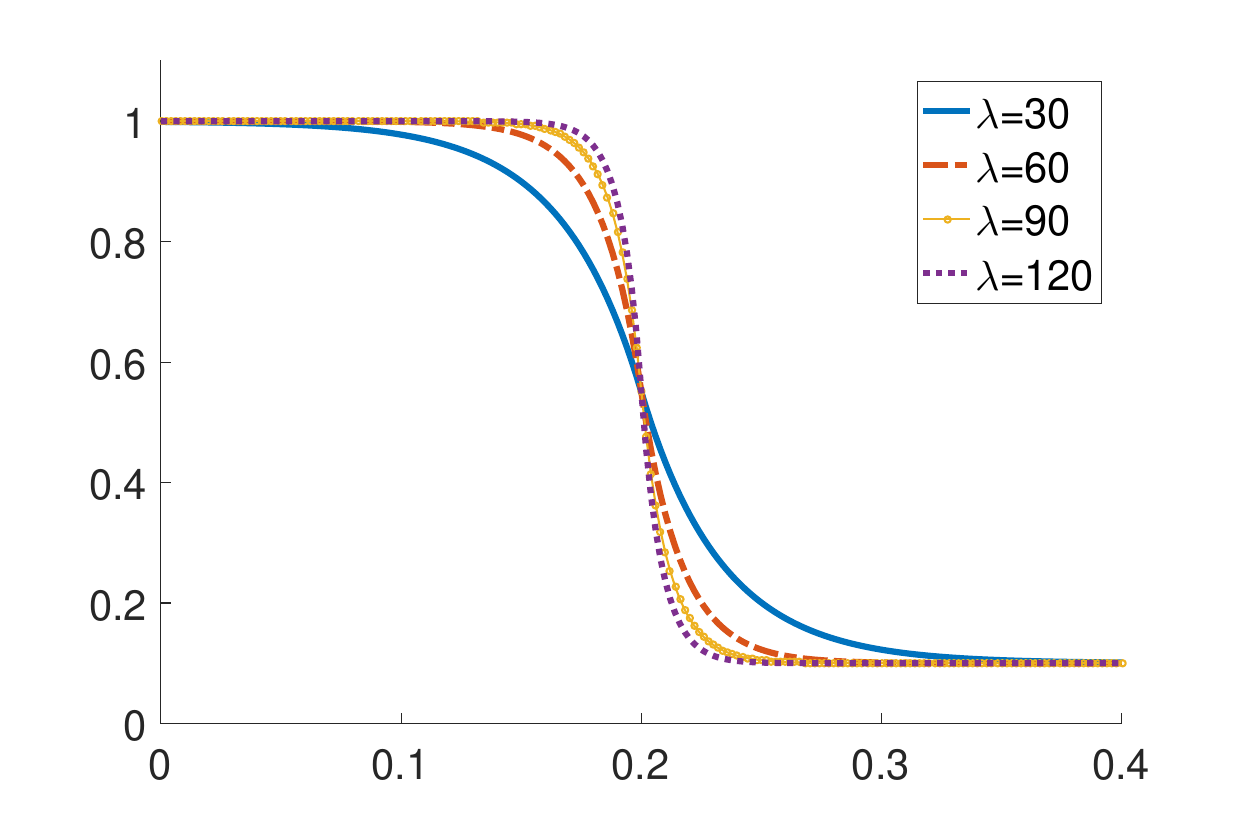}
			\caption{$g(D)$ as described in Example \ref{example: 6}, for $D_0=0.2$ and $\epsilon=0.1$, parametrized by $\lambda$.}
			\label{fig: g soft}
		\end{figure}
		\begin{figure}[htb!]
			\centering
			\includegraphics[width= 0.45 \columnwidth]{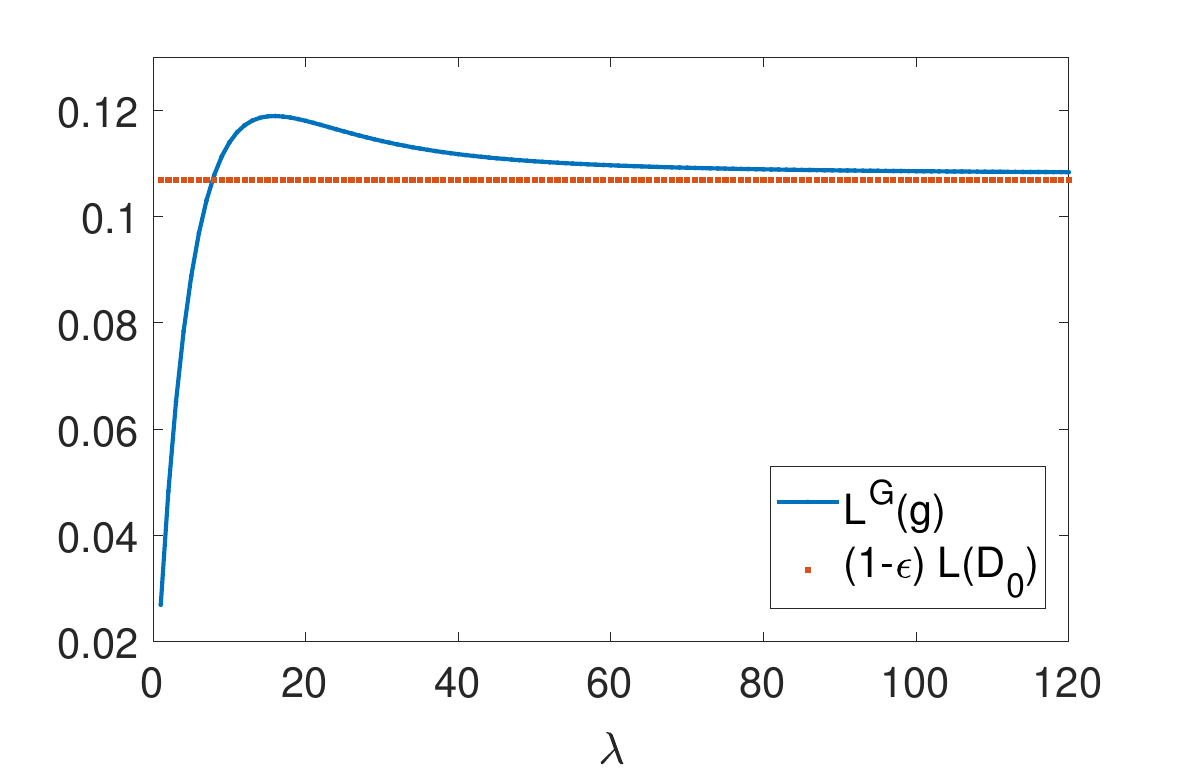}
			\caption{$L^{(G)}(g)$ for the $g$ function given in Example \ref{example: 6}.}
			\label{fig: LGg}
		\end{figure}
		\label{example: 6}
	\end{example}
	
	\section{Proofs \label{sec: proofs}}
	Before proving our main results, we first review Hoeffding's inequality, a version of Chernoff bound used for bounded random variables.
	\begin{lemma} [Hoeffding's inequality {\cite[Theorem~2]{Hoeffding}}]
		Let $X_1, \ldots , X_n$ bounded independent random variables, i.e. $a_i \le X_i \le b_i$ for each $1\le i\le n$. We define the empirical mean of these variables by
		${\displaystyle {\bar{X}}={\frac {1}{n}}(X_{1}+ \ldots +X_{n})}$. Then
		\begin{align}
		{\displaystyle {\begin{aligned}\mathbb {P} \left({\bar{X}}-\mathbb {E} \left[{\bar{X}}\right]\geq t\right)&\leq \exp \left(-{\frac {2n^{2}t^{2}}{\sum _{i=1}^{n}(b_{i}-a_{i})^{2}}}\right),\end{aligned}}}
		\end{align}
		where $t$ is positive, and $E[X]$ is the expected value of $X$.
		\label{lemma: Hoeffding}
	\end{lemma}
	
		\begin{lemma}
			Let $\mu \triangleq P_{X,Y}$ and $\Delta_D \triangleq D_{\max}-D_{\min}$. For any source distribution $\nu \in \mathcal{P}(\mathcal{X}\times\mathcal{Y})$, define
			\begin{equation}
			L_{\nu}(D) \triangleq \min_{P_{\hat{X}|X,Y}: \mathbb{E}_{\nu}[d(X,\hat{X})] \le D} I_{\nu}(Y;\hat{X}).
			\label{eq: source dependent leakage}
			\end{equation}
			Then there exists a non-decreasing concave function $\phi:[0,2]\to[0,\infty)$ with $\phi(0)=0$ and $\lim_{\delta\downarrow0}\phi(\delta)=0$ such that for every $\nu$ and every $D$,
			\begin{equation}
			L_{\nu}(D) \ge L\left(\min\left\{D + \Delta_D\|\nu-\mu\|_1,D_{\max}\right\}\right) - \phi\left(\|\nu-\mu\|_1\right).
			\label{eq: source perturbation lemma}
			\end{equation}
			\label{lemma: source perturbation}
		\end{lemma}
	\begin{IEEEproof}
		Because all alphabets are finite, the set of channels $P_{\hat{X}|X,Y}$ is compact, and the map
		\[
		(\nu,P_{\hat{X}|X,Y}) \mapsto I_{\nu}(Y;\hat{X})
		\]
			is continuous, hence uniformly continuous, on the compact set $\mathcal{P}(\mathcal{X}\times\mathcal{Y}) \times \mathcal{P}(\hat{\mathcal{X}}|\mathcal{X}\times\mathcal{Y})$. Therefore, after replacing the corresponding uniform-continuity function by its concave majorant if necessary, there exists a non-decreasing concave function $\phi$ with the stated properties such that for every $\nu$ and every channel $P_{\hat{X}|X,Y}$,
		\begin{equation}
		I_{\nu}(Y;\hat{X}) \ge I_{\mu}(Y;\hat{X}) - \phi\left(\|\nu-\mu\|_1\right).
		\label{eq: mutual information perturbation}
			\end{equation}
			Moreover, for the same channel,
			\begin{equation}
			\left| \mathbb{E}_{\nu}[d(X,\hat{X})] - \mathbb{E}_{\mu}[d(X,\hat{X})] \right|
			\le \Delta_D\|\nu-\mu\|_1.
			\label{eq: distortion perturbation}
			\end{equation}
			Now let $P^*_{\hat{X}|X,Y}$ achieve $L_{\nu}(D)$. By \eqref{eq: distortion perturbation}, the same channel is feasible for $L\left(D + \Delta_D\|\nu-\mu\|_1\right)$ under the source law $\mu$. Hence,
			\begin{align}
			L_{\nu}(D)
			&= I_{\nu}(Y;\hat{X}) \nonumber\\
			&\ge I_{\mu}(Y;\hat{X}) - \phi\left(\|\nu-\mu\|_1\right) \nonumber\\
			&\ge L\left(\min\left\{D + \Delta_D\|\nu-\mu\|_1,D_{\max}\right\}\right) - \phi\left(\|\nu-\mu\|_1\right),
			\end{align}
			which proves \eqref{eq: source perturbation lemma}.
		\end{IEEEproof}

	\subsection{Proof of Theorem \ref{theorem: f approximation memoryless}} \label{proof: f approximation memoryless}
	Assuming a stationary memoryless mechanism, we provide upper and lower bounds on $\mathbb{E}(f(d(X^n,\hX^n)))$ in terms of $f(\mathbb{E}[d(X^n,\hX^n)]$. This in turn allows us to bound $L^{(M)}(\cdot)$ in terms of $L(f^{-1}(\cdot))$.
	Let $\delta_n=(D_\text{max} -D_\text{min})\sqrt{\log n / n}$. Then, for large enough $n$ we have
	\begin{subequations}
		\label{eq: upper bound on E(d)}
		\allowdisplaybreaks
		\begin{align}
		& \mathbb{E} \left[ f(d(X^n,\hX^n)) \right]\\
		& \le \mathbb{P}\left(d(X^n,\hX^n) \le \mathbb{E}[d(X^n,\hX^n)]  + \delta_n \right) \nonumber\\
		& \quad \cdot f\left(\mathbb{E}[d(X^n,\hX^n)] + \delta_n\right)  \nonumber\\
		& \quad +  \mathbb{P}\left(d(X^n,\hX^n) > \mathbb{E}[d(X^n,\hX^n) ] + \delta_n \right) f(D_{\text{max}})\\
		& \le f(\mathbb{E}[d(X^n,\hX^n)] + \delta_n) + f(D_{\text{max}}) e^{-n \frac{\delta^2_n}{(D_{\text{max}} - D_{\text{min}})^2 }} \label{eq:feasibility of memoryless mech 1}\\
		& \le f\left(\mathbb{E}[d(X^n,\hX^n)] + \delta_n\right) + \frac{f(D_{\text{max}})}{n}, \label{eq:feasibility of memoryless mech 2}
		\end{align}
	\end{subequations}
	where \eqref{eq:feasibility of memoryless mech 1} is due to Lemma \ref{lemma: Hoeffding} and \eqref{eq:feasibility of memoryless mech 2} follows from the definition of $\delta_n$. If $\mathbb{E} \left[ d(X^n,\hX^n) \right] \le f^{-1}_l\left(t-\frac{f(D_{\text{max}})}{n}\right) - \delta_n$, then $\mathbb{E} \left[ f \left (d(X^n,\hX^n)\right ) \right] \le t$, and we have
	\begin{equation}
	L^{(M)} (n,t,f) \le L\left(f^{-1}_l\left(t-\frac{f(D_{\text{max}})}{n}\right) - \delta_n\right) \label{eq: LM upper}.
	\end{equation}
	Since $f^{-1}_l(\cdot)$ is left-continuous, and $L(\cdot)$ is continuous, taking the limit as $n \rightarrow \infty$ gives
	\begin{equation}
	L^{(M)} (t,f) \le L\left(f^{-1}_l\left(t\right) \right) \label{eq: LM upper limit}.
	\end{equation}
	With a similar argument and using the negative of the distortion function in Lemma \ref{lemma: Hoeffding}, we have
	\begin{subequations}
		\label{eq: lower bound on E(d)}
		\allowdisplaybreaks
		\begin{align}
		& \mathbb{E} \left[ f(d(X^n,\hX^n)) \right]\\
		& \ge \mathbb{P}\left(d(X^n,\hX^n) \ge \mathbb{E}[d(X^n,\hX^n)]  - \delta_n \right) \nonumber\\
		& \quad \cdot f\left(\mathbb{E}[d(X^n,\hX^n)] - \delta_n\right)  \nonumber\\
		& \quad +  \mathbb{P}\left(d(X^n,\hX^n) < \mathbb{E}[d(X^n,\hX^n) ] - \delta_n \right) f(D_{\text{min}})\\
		& \ge \left(1 - \mathbb{P}\left(d(X^n,\hX^n) < \mathbb{E}[d(X^n,\hX^n) ] - \delta_n \right)\right)\nonumber\\
		&\quad \cdot f(\mathbb{E}[d(X^n,\hX^n)] - \delta_n) \\
		& \ge \left(1- \frac{1}{n}\right)  f\left(\mathbb{E}[d(X^n,\hX^n)] - \delta_n\right) \label{eq: converse of memoryless mech 1},
		\end{align}
	\end{subequations}
	where \eqref{eq: converse of memoryless mech 1} is due to Lemma \ref{lemma: Hoeffding}.
	Therefore, if 
	\begin{equation}
	\mathbb{E} \left[ f \left (d(X^n,\hX^n)\right ) \right] \le t,
	\end{equation}
	then
	\begin{equation}
	\mathbb{E} \left[ d(X^n,\hX^n) \right] \le f^{-1}_u\left(t\left(1+\frac{1}{n-1}\right)\right)+\delta_n,
	\end{equation}
	and we have
	\begin{equation}
	L\left(f^{-1}_u\left(t\left(1+\frac{1}{n-1}\right)\right)+\delta_n\right) \le L^{(M)} (n,t,f). \label{eq: LM lower} 
	\end{equation}
	Since $f^{-1}_u(\cdot)$ is right-continuous, and $L(\cdot)$ is continuous, taking the limit as $n \rightarrow \infty$ gives
	\begin{equation}
	L\left(f^{-1}_u(t)\right) \le L^{(M)} (t,f).  \label{eq: LM lower limit} 
	\end{equation}
	
	Recall the definition of $\mc{T}_f$ in \eqref{eq: T_f definition}. If $t \notin \mc{T}_f$, then $f^{-1}_l(t)=f^{-1}_u(t)$. Hence, \eqref{eq: LM upper limit} and \eqref{eq: LM lower limit} imply \eqref{eq: first order leakage memoryless 1}. Otherwise, if $t \in \mc{T}_f$, the same two bounds directly yield \eqref{eq: first order leakage memoryless 2}. The exceptional set $\mc{T}_f$ is countable, and the sharper behavior at those isolated thresholds is not needed in the proof of Theorem \ref{theorem: f approximation}.

	\subsection{Proof of Lemma \ref{lemma: g bounded asymptotic leakage - simple}} \label{proof: g bounded asymptotic leakage - simple}
		\textbf{Achievability:}
		We build a combination of memoryless mechanisms to show achievability. Specifically, we pick the optimal mechanisms for single letter leakage functions evaluated at approximately $D_1, D_2, \ldots, D_k$. The reason for not choosing the exact values of $D_i$ is that we need the optimal single letter mechanism to satisfy a slightly smaller average distortion bound so that a tail probability constraint is guaranteed.
		
		Recall that $\mathcal{P}^*(D)$ is the set of optimal single letter mechanisms for $L(D)$.
		Let $\Delta_D \triangleq D_{\max}-D_{\min}$.
		Then, for any $D$ let $P^{*(D)}_{\hat{X}|X,Y}\in \mathcal{P}^*(D)$ and $P^{*(D)}_{\hat{X}^n|X^n,Y^n}=\left(P^{*(D)}_{\hat{X}|X,Y}\right)^n$. Define $\epsilon^{(n)}_0 = \epsilon_0 = 1$, and for any $1\le i \le k$ let
		\begin{align}
		D^{(n)}_i &\triangleq D_i - \Delta_D \sqrt{\frac{\log n}{n}} \label{eq: D^n},\\
		\epsilon^{(n)}_i &\triangleq \frac{\epsilon_i -  \frac{\epsilon_{i-1}}{n}  - e^{-n\delta(i)} }{1- \frac{1}{n}} \label{eq: eps^n}.
		\end{align}
		For the special case where $D_1=D_{\text{min}}$, let $D^{(n)}_1 = D_1$ instead.
		Note that for sufficiently large $n$ we have $D_{\min} \le D^{(n)}_i \le D_i$ and $0 \le \epsilon^{(n)}_i \le \epsilon_i$, which implies that
		\begin{equation}
		\epsilon^{(n)}_i = \epsilon_i + O\left(\frac{1}{n}\right) \label{eq: epsilon and epsilon^n}.
		\end{equation}
		Now let $E$ be a random variable independent from $(X^n,Y^n)$ with alphabet set $\{1,\ldots,k+1\}$, where $P(E=i) = \epsilon^{(n)}_{i-1} - \epsilon^{(n)}_{i}$ for $1 \le i \le k$, and $P(E=k+1) = \epsilon^{(n)}_{k}$. Then, consider the following mechanism:
	\begin{align}
	&P_{\hX^n|X^n,Y^n}(\hx^n|x^n,y^n) \nonumber\\
	&= \begin{cases}
	P^{*(D^{(n)}_{E})}_{\hX^n|X^n,Y^n}(\hx^n|x^n,y^n), & \text{if } \quad 1\le E \le k,\\
	P^{*(D^{(n)}_k)}_{\hX^n} (\hx^n), & \text{if } \quad E=k+1.
	\end{cases}
	\label{eq: P achievability}
	\end{align}
		First, we show that it is feasible, i.e. it satisfies $P(d(X^n,\hX^n) > D_i) \le \epsilon_i$ for any $ 1 \le i \le k$. Since $D^{(n)}_i \rightarrow D_i$, and $D_i$ has a distinct value for each $i$, there exists a $\delta(i) > 0$ and $n_i$ such that $\delta(i) < e^{- \frac{\left(D_i - D^{(n)}_{i-1}\right)^2}{\Delta_D^2}}$ for $n \ge n_i$. Therefore, for any $1 \le i \le k$ and $n \ge n_i$ we can bound the $i$th error probability by
	\begin{subequations}
		\begingroup
		\allowdisplaybreaks
		\begin{align}
		&\mathbb{P}[d(X^n,\hX^n) >D_i] \nonumber\\
		&= \epsilon^{(n)}_{k+1} P(d(X^n,\hX^n) >D_i | E=k+1) \nonumber\\
		& \quad + \sum_{j=1}^{k}  \left( \epsilon^{(n)}_{j-1} - \epsilon^{(n)}_j  \right) P(d(X^n,\hX^n) >D_i | E=j)\nonumber \\
		& \le \epsilon^{(n)}_i  + \sum_{j=1}^{i}  \left( \epsilon^{(n)}_{j-1} - \epsilon^{(n)}_j  \right) P(d(X^n,\hX^n) >D_i | E=j)\\
		&\le \epsilon^{(n)}_i  + \sum_{j=1}^{i}  \left( \epsilon^{(n)}_{j-1} - \epsilon^{(n)}_j  \right) e^{-n \frac{\left(D_i - D^{(n)}_j\right)^2}{\Delta_D^2}} \label{eq: Chernoff bound}\\
		&\le \epsilon^{(n)}_i + e^{-n \frac{\left(D_i - D^{(n)}_{i-1}\right)^2}{\Delta_D^2}}  \sum_{j=1}^{i-1}  \left( \epsilon^{(n)}_{j-1} - \epsilon^{(n)}_j  \right) \nonumber \\
		& \quad + \left( \epsilon^{(n)}_{i-1} - \epsilon^{(n)}_i  \right) e^{-n \frac{\left(D_i - D^{(n)}_i\right)^2}{\Delta_D^2}}\\
		& = \epsilon^{(n)}_i + e^{-n \frac{\left(D_i - D^{(n)}_{i-1}\right)^2}{\Delta_D^2}}  \left( 1 - \epsilon^{(n)}_{i-1}  \right) + \frac{1}{n}\left( \epsilon^{(n)}_{i-1} - \epsilon^{(n)}_i  \right) \\
		& \le \epsilon^{(n)}_i + e^{-n\delta(i)}   \left( 1 - \epsilon^{(n)}_{i-1}  \right) + \frac{1}{n}\left( \epsilon^{(n)}_{i-1} - \epsilon^{(n)}_i  \right)  \label{eq: delta definition}\\
		& \le  \epsilon^{(n)}_i + \frac{1}{n}\left( \epsilon^{(n)}_{i-1} - \epsilon^{(n)}_i  \right)  + e^{-n\delta(i)} \\
		& \le  \epsilon^{(n)}_i + \frac{1}{n}\left( \epsilon_{i-1} - \epsilon^{(n)}_i  \right)  + e^{-n\delta(i)} \\
		& \le \epsilon_i \label{eq: total probability of error is less than epsilon},
		\end{align}
		\endgroup
	\end{subequations}
	where \eqref{eq: Chernoff bound} follows from Lemma \ref{lemma: Hoeffding}, \eqref{eq: delta definition} is due to the definition of $\delta(i)$, and \eqref{eq: total probability of error is less than epsilon} results from \eqref{eq: D^n} and \eqref{eq: eps^n}.
	Note that in the special case where $D_1=D_{\text{min}}$, we have $D^{(n)}_1 = D_1=D_{\text{min}}$. Therefore, $\mathbb{P}[d(X^n,\hX^n) >D_1] = 0$, because the optimal mechanism achieving $L(D_{\text{min}})$ has to satisfy $\mathbb{P}[d(X^n,\hX^n) = D_1] = 1$.
	
	We now show that the mechanism introduced in \eqref{eq: P achievability} achieves \eqref{eq: multi constraint}. Recalling the definition of $E$ we have
	\begin{subequations}
		\begingroup
		\allowdisplaybreaks
		\begin{align}
		&I(Y^n;\hX^n)  \\
		&\le I(Y^n;\hX^n|E) \\
		&= \sum_{j=1}^{k} (\epsilon^{(n)}_{j-1}-\epsilon^{(n)}_{j}) I(Y^n;\hX^n|E=j) \nonumber\\
		& \quad + \epsilon^{(n)}_{k} \; I(Y^n;\hX^n|E=k+1)\\
		&=\sum_{j=1}^{k} (\epsilon^{(n)}_{j-1}-\epsilon^{(n)}_{j}) n \; L(D^{(n)}_j) \label{eq: lemma 1 UB}\\
		& = n \sum_{j=1}^{k} (\epsilon_{j-1}-\epsilon_{j}) \; L(D_j)  + O(\sqrt{n\log n}),
		\label{eq: lemma 1 UB part2}
		\end{align}
		\endgroup
	\end{subequations}
	where \eqref{eq: lemma 1 UB} is due to definition of the chosen mechanism in \eqref{eq: P achievability}, and \eqref{eq: lemma 1 UB part2} is implied by \eqref{eq: D^n}. This yields the upper bound in \eqref{eq: multi constraint LB and UB}.
	
	
	\textbf{Converse:}
	Assume a mechanism $P_{\hat{X}^n|X^n,Y^n}$ satisfying the feasibility constraint of \eqref{eq: multi constraint}. Define the indicator random variable $E$ as
	\begin{equation}
	E = \begin{cases}
	1, & \text{if } d(X^n,\hX^n) \le D_1,\\
	2, & \text{if } D_1 < d(X^n,\hX^n) \le D_2,\\
	\vdots&\vdots\\
	k+1, & \text{if } D_k < d(X^n,\hX^n).
	\end{cases}
	\end{equation}
	Let $p_i \triangleq \mathbb{P}[E=i]$, let $D_{k+1}\triangleq D_{\max}$, and let $\mu \triangleq P_{X,Y}$. Since $E$ takes values in an alphabet of size $k+1$, we have $H(E)\le \log(k+1)$. Also, for each $1\le i\le k$,
	\[
	\mathbb{E}[d(X^n,\hat{X}^n)\mid E=i]\le D_i,
	\]
	while the same statement is trivially true for $i=k+1$ with $D_{k+1}=D_{\max}$.
	
	Because $I(Y^n;E\mid \hat{X}^n)\le H(E)$, we can lower bound the leakage as
	\begin{subequations}
		\begingroup
		\allowdisplaybreaks
		\begin{align}
		I(Y^n;\hat{X}^n)
		&\ge I(Y^n;\hat{X}^n,E)-H(E) \nonumber\\
		&= \sum_{j=1}^{n} I(Y_j;\hat{X}^n,E\mid Y^{j-1})-H(E) \nonumber\\
		&\ge \sum_{j=1}^{n} I(Y_j;\hat{X}_j,E)-H(E) \nonumber\\
		&\ge \sum_{i=1}^{k+1} p_i \sum_{j=1}^{n} I(Y_j;\hat{X}_j\mid E=i)-H(E).
		\label{eq: converse of Lemma 1}
		\end{align}
		\endgroup
	\end{subequations}
	
			Let $\Delta_D \triangleq D_{\max}-D_{\min}$, and let $\omega:[0,\Delta_D] \to [0,\infty)$ be a non-decreasing concave function such that $\omega(0)=0$ and
			\[
			|L(D)-L(D')| \le \omega(|D-D'|), \qquad D,D'\in[D_{\min},D_{\max}].
			\]
			Define
			\begin{equation}
			\psi(u) \triangleq \omega\left(\min\left\{\Delta_D u,\Delta_D\right\}\right) + \phi(u), \qquad 0\le u\le 2.
			\label{eq: psi definition}
			\end{equation}
			Then $\psi$ is non-decreasing, concave, and satisfies $\psi(0)=0$.
			For each $i$ with $p_i>0$ and each $j\in\{1,\ldots,n\}$, define
			\[
			\nu_{ij} \triangleq P_{X_j,Y_j\mid E=i}, \qquad t_{ij} \triangleq \mathbb{E}[d(X_j,\hat{X}_j)\mid E=i], \qquad a_{ij} \triangleq \|\nu_{ij}-\mu\|_1.
		\]
		Since
		\[
		P_{X^n,Y^n\mid E=i}(x^n,y^n) = \frac{P(E=i\mid x^n,y^n)}{p_i} \mu^{\otimes n}(x^n,y^n) \le \frac{1}{p_i}\mu^{\otimes n}(x^n,y^n),
	\]
	we have
		\[
		D(P_{X^n,Y^n\mid E=i}\|\mu^{\otimes n}) \le \log \frac{1}{p_i}.
		\]
		By the chain rule for KL divergence, convexity of KL, and Pinsker's inequality,
		\begin{equation}
		\sum_{j=1}^{n} a_{ij}^2
		\le 2 \sum_{j=1}^{n} D(\nu_{ij}\|\mu)
		\le 2 \log \frac{1}{p_i}.
		\label{eq: total error bound on distances}
		\end{equation}
		Define
		\begin{equation}
		\bar{a}_i \triangleq \frac{1}{n}\sum_{j=1}^{n} a_{ij}.
		\label{eq: average perturbation}
		\end{equation}
		Then Jensen's inequality and \eqref{eq: total error bound on distances} yield
		\begin{equation}
		\bar{a}_i \le \sqrt{\frac{1}{n}\sum_{j=1}^{n} a_{ij}^2} \le \sqrt{\frac{2\log(1/p_i)}{n}}.
		\label{eq: average perturbation bound}
		\end{equation}
		
		Now fix $i$ and $j$. By the definition of $L_{\nu}$ in \eqref{eq: source dependent leakage},
		\[
			I(Y_j;\hat{X}_j\mid E=i) \ge L_{\nu_{ij}}(t_{ij}).
			\]
			Hence, by Lemma \ref{lemma: source perturbation},
			\begin{equation}
			I(Y_j;\hat{X}_j\mid E=i) \ge L\left(\min\left\{t_{ij}+\Delta_D a_{ij},D_{\max}\right\}\right)-\phi(a_{ij}),
			\label{eq: L is a lower bound on conditioned I}
			\end{equation}
			Define
			\[
			z_{ij} \triangleq \min\left\{t_{ij}+\Delta_D a_{ij},D_{\max}\right\}.
			\]
			Then
			\begin{subequations}
				\begingroup
				\allowdisplaybreaks
			\begin{align}
				\frac{1}{n}\sum_{j=1}^{n} I(Y_j;\hat{X}_j\mid E=i)
				&\ge \frac{1}{n}\sum_{j=1}^{n} L(z_{ij}) - \frac{1}{n}\sum_{j=1}^{n}\phi(a_{ij}) \nonumber\\
				&\ge L\left(\frac{1}{n}\sum_{j=1}^{n} z_{ij}\right) - \phi(\bar{a}_i) \nonumber\\
				&\ge L\left(\min\left\{D_i + \Delta_D\bar{a}_i,D_{\max}\right\}\right) - \phi(\bar{a}_i) \nonumber\\
				&\ge L(D_i) - \psi(\bar{a}_i),
				\label{eq: lemma 1 LB special}
				\end{align}
				\endgroup
			\end{subequations}
			where the second step uses convexity of $L$ together with concavity of $\phi$, and the third uses
			\[
			\frac{1}{n}\sum_{j=1}^{n} z_{ij}
			\le \frac{1}{n}\sum_{j=1}^{n} t_{ij} + \Delta_D\bar{a}_i
			\le D_i + \Delta_D\bar{a}_i.
			\]
			
			Combining \eqref{eq: converse of Lemma 1} and \eqref{eq: lemma 1 LB special}, and using concavity of $\psi$, we obtain
			\begin{subequations}
				\begingroup
				\allowdisplaybreaks
				\begin{align}
				\frac{1}{n}I(Y^n;\hat{X}^n)
				&\ge \sum_{i=1}^{k+1} p_i L(D_i) - \sum_{i=1}^{k+1} p_i \psi(\bar{a}_i) - \frac{H(E)}{n} \nonumber\\
				&\ge \sum_{i=1}^{k+1} p_i L(D_i) - \psi\left(\sum_{i=1}^{k+1} p_i \bar{a}_i\right) - \frac{H(E)}{n}.
				\label{eq: lemma 1 LB before epsilon}
				\end{align}
				\endgroup
			\end{subequations}
		Using \eqref{eq: average perturbation bound}, the definition of $H(E)$, and concavity of the square-root function gives
		\[
		\sum_{i=1}^{k+1} p_i \bar{a}_i
		\le \sqrt{\frac{2}{n}} \sum_{i=1}^{k+1} p_i \sqrt{\log \frac{1}{p_i}}
		\le \sqrt{\frac{2H(E)}{n}}.
		\]
			Therefore,
			\begin{equation}
			\frac{1}{n}I(Y^n;\hat{X}^n)
			\ge \sum_{i=1}^{k+1} p_i L(D_i) - \psi\left(\sqrt{\frac{2H(E)}{n}}\right) - \frac{H(E)}{n}.
			\label{eq: lemma 1 LB with delta}
			\end{equation}
			Since $H(E)\le \log(k+1)$ and $\psi$ is non-decreasing,
			\begin{equation}
			\frac{1}{n}I(Y^n;\hat{X}^n)
			\ge \sum_{i=1}^{k+1} p_i L(D_i) - \psi\left(\sqrt{\frac{2\log(k+1)}{n}}\right) - \frac{\log(k+1)}{n}.
			\label{eq: lemma 1 LB with explicit beta}
			\end{equation}
		
		Now let $q_i \triangleq \mathbb{P}[E\ge i]$ for $1\le i\le k+1$, and set $q_{k+2}=0$. Then $q_1=1$, $q_{i+1}=\mathbb{P}[d(X^n,\hat{X}^n)>D_i]\le \epsilon_i$ for $1\le i\le k$, and $p_i=q_i-q_{i+1}$. Since $L(D_i)$ is non-increasing in $i$ and $L(D_{k+1})=L(D_{\max})=0$, we have
		\begin{align}
		\sum_{i=1}^{k+1} p_i L(D_i)
	&= \sum_{i=1}^{k+1} (q_i-q_{i+1})L(D_i) \nonumber\\
	&= L(D_1) + \sum_{i=2}^{k+1} q_i \left(L(D_i)-L(D_{i-1})\right) \nonumber\\
	&\ge L(D_1) + \sum_{i=2}^{k+1} \epsilon_{i-1}\left(L(D_i)-L(D_{i-1})\right) \nonumber\\
	&= \sum_{i=1}^{k} (\epsilon_{i-1}-\epsilon_i)L(D_i).
	\label{eq: p_i and eps_i}
	\end{align}
			Since $\psi(0)=0$, the quantity $\beta_n$ defined in \eqref{eq: beta_n definition} satisfies $\beta_n \to 0$. Combining \eqref{eq: lemma 1 LB with explicit beta} and \eqref{eq: p_i and eps_i} yields
			\[
			\frac{1}{n}I(Y^n;\hat{X}^n)
			\ge \sum_{i=1}^{k} (\epsilon_{i-1}-\epsilon_i)L(D_i) - \beta_n,
			\]
			which establishes the lower bound in \eqref{eq: multi constraint LB and UB}.

	\subsection{Proof of Theorem \ref{theorem: f approximation}} \label{proof: f approximation}
	We will need the following lemma in our proof for Theorem \ref{theorem: f approximation}.
	\begin{lemma}
		For any given $n$ and $f$, $L^{(G)} (n,t,f)$ is convex in $t$. Consequently, $L^{(G)} (t,f)$ is also convex in $t$, for any $f$.
		\label{lemma: LG is convex in t}
	\end{lemma}
	\begin{IEEEproof}
		For any $t_1,t_2$, and some $0\le\lambda \le 1$, let $t_{\lambda} = \lambda t_1 + (1-\lambda) t_2$. We will show that $L(n,t_{\lambda},f) \le \lambda L(n,t_1,f) + (1-\lambda) L(n,t_2,f)$. Let $P_1$ and $P_2$ be optimal mechanisms for $L^{(G)}(n,t_1,f)$ and $L^{(G)}(n,t_2,f)$ respectively, and $P_{\lambda} \triangleq \lambda P_1 + (1-\lambda) P_2$. Note that $P_\lambda$ is feasible for $L^{(G)}(n,t_\lambda,f)$ because 
		\begin{align}
		&\mathbb{E}_{P_{\lambda}}\left[f\left(d(X^n,\hX^n)\right)\right] \nonumber \\
		&= \lambda \mathbb{E}_{P_{1}}\left[f\left(d(X^n,\hX^n)\right)\right] + (1-\lambda) \mathbb{E}_{P_2}\left[f(d(X^n,\hX^n))\right] \nonumber\\
		&\le \lambda t_1 + (1-\lambda) t_2 = t_{\lambda}.
		\end{align}
		Moreover, since $I(Y^n;\hX^n)$ is convex in $P_{\hX^n|X^n,Y^n}$, the leakage achieved by $P_\lambda$ is at most equal to $\lambda L^{(G)}(n,t_1,f) + (1-\lambda) L^{(G)}(n,t_2,f)$ which implies $L^{(G)}(n,t_{\lambda},f) \le \lambda L^{(G)}(n,t_1,f) + (1-\lambda) L^{(G)}(n,t_2,f)$. Finally we note that the asymptotic leakage $L^{(G)}(t,f)$ is also convex in $t$ because it is the limit of convex functions in $t$.
	\end{IEEEproof}
	
	We now present an achievable scheme and a converse for Theorem \ref{theorem: f approximation}.
	
	\textbf{Achievability:}
	We know that $L^{(G)}(t,f) \le L^{(M)}(t,f) \le L(f^{-1}_l(t))$, where the latter inequality is due to Theorem \ref{theorem: f approximation memoryless}. Since by Lemma \ref{lemma: LG is convex in t}, $L^{(G)}(t,f)$ is a convex function in $t$, the definition of lower convex envelope gives $L^{(G)}(t,f) \le (L \circ f^{-1}_l)^{**}(t)$. This in turn gives $L^{(G)}(t,f) \le (L \circ f^{-1})^{**} (t)$ due to Remark \ref{remark: upper and lower g** are equal}.
	
	\textbf{Converse:}
	We first focus on the class of piecewise step functions $f$, and then show that the result holds for any function $f$, using piecewise step approximations of $f$.

	\textit{Piecewise Step functions $f$:} Let us consider the class of functions $f$ that are of the form
	\begin{equation}
	f(D) = \sum_{i=1}^{k} a_i \boldsymbol{1}(D>D_i),
	\label{eq: f approx}
	\end{equation}
	where $k$ is finite and each $D_i$ is a distinct distortion level with $f(D_i) < f(D_j)$ for $i<j$. For this class of functions, \eqref{eq: LG definition} simplifies and can be lower bounded as
	\begin{subequations}
		\begingroup
		\allowdisplaybreaks
		\begin{align}
		&L^{(G)}(n,t,f) = \min_{\substack{P_{\hX^n|X^n,Y^n} :\\ \mathbb{E}[\sum_{i=1}^{k} a_i \boldsymbol{1} (d(X^n,\hX^n) > D_i)]\le t}} \frac{1}{n}I(Y^n;\hat{X}^n) \nonumber\\
		& =\min_{\substack{P_{\hX^n|X^n,Y^n} :\\ \sum_{i=1}^{k} a_i \mathbb{P}[ (d(X^n,\hX^n) > D_i)]\le t}} \frac{1}{n}I(Y^n;\hat{X}^n)\\
		&= \min_{\substack{0 \le \epsilon_k\le\ldots\le\epsilon_1\le 1:\\ \sum_{i=1}^{k} a_i \epsilon_i\le t}} \min_{\substack{P_{\hX^n|X^n,Y^n} :\\ \mathbb{P}[d(X^n,\hX^n) > D_i] \le \epsilon_i, \\ \forall 1 \le i  \le k}} \frac{1}{n}I(Y^n;\hat{X}^n)\\
		&\ge \min_{\substack{0 \le \epsilon_k\le\ldots\le\epsilon_1\le 1:\\ \sum_{i=1}^{k} a_i \epsilon_i\le t}} \sum_{i=1}^{k} (\epsilon_{i-1} - \epsilon_{i})  L(D_i) \nonumber\\
		& \qquad + \epsilon_k L(D_{\text{max}}) - \beta_n \label{eq: optimization converse 1}\\
		&\ge \max_{\lambda \ge 0} \min_{0 \le \epsilon_k\le\ldots\le\epsilon_1\le 1} \sum_{i=1}^{k} (\epsilon_{i-1} - \epsilon_{i}) L(D_i) + \epsilon_k L(D_{\text{max}}) \nonumber\\
		&\qquad  +\lambda \sum_{i=1}^{k} a_i \epsilon_i - \lambda t - \beta_n\label{eq: optimization converse 2}\\
		&= \max_{\lambda \ge 0} \min_{\substack{ \gamma_1 \cdots \gamma_{k+1} :\\ \gamma_i \ge 0, \forall i=1,\ldots,k+1,\\\sum_{i=1}^{k+1} \gamma_i=1}} \sum_{i=1}^{k+1} \gamma_i L(D_i) - \beta_n \nonumber\\
		& \qquad +\lambda \sum_{i=1}^{k+1} f(D_i)\gamma_i - \lambda t \label{eq: optimization converse 3}\\
		& = \max_{\lambda \ge 0}  \min_i L(D_i)+\lambda f(D_i)-\lambda t - \beta_n \label{eq: optimization converse 4}\\
		& = \max_{\lambda \ge 0}  \min_i L(f^{-1}_u(t_i))+\lambda t_i-\lambda t - \beta_n\label{eq: optimization converse 5}
		\end{align}
		\endgroup
	\end{subequations}
	where
	\begin{itemize}
		\item \eqref{eq: optimization converse 1} follows from Lemma \ref{lemma: g bounded asymptotic leakage - simple}, with the corresponding vanishing sequence $\beta_n$, and the fact that $L(D_{\text{max}})=0$,
		\item \eqref{eq: optimization converse 2} is due to forming the Lagrangian given by incorporating only the last constraint in \eqref{eq: optimization converse 1}, i.e. $\sum_{i=1}^{k} a_i \epsilon_i \le t$,
		\item \eqref{eq: optimization converse 3} is derived by letting $\epsilon_{k+1} = 0$, $D_{k+1} = D_{\text{max}}$, and $\gamma_i = \epsilon_{i-1} - \epsilon_i$, for $i=1,\ldots,k+1$.
		\item \eqref{eq: optimization converse 4} holds because a convex combination of non-negative real numbers is minimized by choosing a $\boldsymbol{\gamma}$ with $\gamma_i=1$ for some $i$ corresponding to the smallest $L(D_i)+\lambda f(D_i)$, and $\gamma_j=0$, for all other $j \neq i$,
		\item and \eqref{eq: optimization converse 5} is derived by defining $t_i=f(D_i)$, i.e. $D_i = f^{-1}_u(t_i)$.
	\end{itemize}
	Then, by taking the limit as $n \rightarrow \infty$ we have
	\begin{equation}
	L^{(G)}(t,f) = \max_{\lambda}  \min_i L(f^{-1}_u(t_i))+\lambda t_i-\lambda t .
	\end{equation}
	Note that the $i$th function, $L(f^{-1}_u(t_i))+\lambda t_i$ is a minimizer for some $\lambda$, if for all $j \neq i$ we have
	\begin{equation}
	L(f^{-1}_u(t_i)) + \lambda t_i \le L(f^{-1}_u(t_j)) + \lambda t_j,
	\end{equation}
	or equivalently
	\begin{subequations}
		\begingroup
		\allowdisplaybreaks
		\begin{align}
		&\frac{L(f^{-1}_u(t_i))-L(f^{-1}_u(t_j))}{t_{i}-t_{j}} \le -\lambda, &\text{   for $j < i$}, \label{eq: on the convex envelope a}\\
		&\frac{L(f^{-1}_u(t_i))-L(f^{-1}_u(t_j))}{t_{i}-t_{j}} \ge -\lambda, &\text{   for $j > i$}\label{eq: on the convex envelope b}.
		\end{align}
		\endgroup
	\end{subequations}
	Note that \eqref{eq: on the convex envelope a} and \eqref{eq: on the convex envelope a} imply the slope of the line connecting points $\{(t_i,L(f^{-1}_u(t_i))), (t_j,L(f^{-1}_u(t_j)))\}$ is not larger than $-\lambda$, for $j < i$, and not smaller than $-\lambda$, for $j >i$. This holds if and only if $L(f^{-1}_u(t_i))= (L \circ f^{-1}_u)^{**}(t_i)$. Since $(L \circ f^{-1}_u)^{**}(t_i) = (L \circ f^{-1})^{**}(t_i)$ due to Remark \ref{remark: upper and lower g** are equal}, the only relevant $i$ in the minimization in \eqref{eq: optimization converse 4} are those for which $L(f^{-1}_u(t_i))=(L \circ f^{-1})^{**}(t_i)$. Hence, \eqref{eq: optimization converse 4} can be rewritten as
	\begin{align}
	&L^{(G)}(t,f)\ge \nonumber\\
	&\max_{\lambda}  \min_{i: L(f^{-1}_u(t_i))=(L \circ f^{-1})^{**}(t_i)} L(f^{-1}_u(t_i))+\lambda t_i-\lambda t.
	\end{align}
	For a chosen $\lambda$ and $i$, $L(f^{-1}_u(t_i))+\lambda t_i-\lambda t$ is the evaluation of a linear function at $t$, which is tangential to $(L \circ f^{-1})^{**}(\cdot)$ at $(t_i, (L \circ f^{-1})^{**}(t_i))$, with slope $-\lambda$. This value is always smaller than or equal to $(L \circ f^{-1})^{**}(t)$, and because $(L \circ f^{-1})^{**}(\cdot)$ is a convex piecewise linear function, it suffices to optimize over only those values of $\lambda$ that are equal to the slope of the linear segment of $(L \circ f^{-1})^{**}(\cdot)$ that contains $t$. Thus, for an optimal $\lambda$ we have $\min_i (L \circ f^{-1}_u)(t_i)+\lambda t_i-\lambda t = (L \circ f^{-1})^{**}(t)$, resulting in $L^{(G)}(t,f) \ge (L \circ f^{-1})^{**}(t)$.
	
	\textit{General functions $f$:} Finally, we now show that $L^{(G)} (t,f) \ge (L \circ f^{-1})^{**}(t)$ for the case of general non-decreasing left continuous functions $f$.
	For any $\delta > 0$, there exists a lower approximation $f_\delta$ of $f$ over $[D_{\text{min}},D_{\text{max}}]$ that has the form of \eqref{eq: f approx} with a finite number of step functions, i.e. $f_\delta(x) = \sum_{i=1}^{k} a_i 1_{D_i}(x)$, with $a_i = f(D_i)-f(D_{i-1})$ for $1\le i \le k$ and $a_{\text{max}} \triangleq \max_{i} a_i \le \delta$. Then, we have $f_\delta(D) < f(D) \le f_\delta(D) + \delta$, and thus
	\begin{subequations}
		\begingroup
		\allowdisplaybreaks
		\begin{align}
		L^{(G)}(t,f) &\ge L^{(G)}(t,f_\delta)\label{eq: lower LG(t,f_delta) le LG(t,f)}\\
		&\ge (L \circ {f_\delta}^{-1}_u)^{**}(t) \label{eq: for f_delta LG ge g**}\\
		& = (L \circ {f_\delta}^{-1})^{**}(t), \label{eq: g**(bar_f_delta) = g**(f_delta)}
		\end{align}
		\endgroup
	\end{subequations}
	where \eqref{eq: lower LG(t,f_delta) le LG(t,f)} holds because we have $L^{(G)}(n,t,f_\delta) \le L^{(G)}(n,t,f)$ for any $n$, \eqref{eq: for f_delta LG ge g**} is based on the result we had earlier on piecewise step functions specifically, and \eqref{eq: g**(bar_f_delta) = g**(f_delta)} is due to Remark \ref{remark: upper and lower g** are equal}.
	Then, taking the limit as $\delta \rightarrow 0$ and the fact that $\lim_{\delta \rightarrow 0} f_\delta(D) = f(D)$ gives $L^{(G)}(t,f) \ge (L \circ f^{-1})^{**}(t)$.

	\subsection{Proof of Theorem \ref{theorem: g bounded asymptotic leakage}} \label{proof: g bounded asymptotic leakage}

	We now proceed to proving the result in \eqref{eq: g bounded asymptotic leakage result} for all non-increasing right-continuous functions $g:[D_{\text{min}},D_{\text{max}}] \rightarrow (0,1]$. Recall that we proved this for simple functions through Lemma \ref{lemma: g bounded asymptotic leakage - simple}. For any bounded, non-increasing, and right-continuous function $g$, there exist two sequences of simple functions $\{\overline{g}_i\}_{i=1}^{\infty}$ and $\{\underline{g}_i\}_{i=1}^{\infty}$ that are bounded away from zero, converge to $g$ uniformly from above and below, respectively, and each of functions $\overline{g}_i$ and $\underline{g}_i$ takes $i$ distinct values. Since $\underline{g}_i(D) \le g(D) \le \overline{g}_i (D)$ for all $i\ge1$, $D \in [D_{\text{min}},D_{\text{max}}]$, and the asymptotic optimal leakage for simple constraint functions is the integral in \eqref{eq: multi constraint result}, for each $i \ge 1$ we have
	\begin{equation}
	\int_{D_{\text{min}}}^{D_{\text{max}}} L(D) d(\overline{g}_i(D)) \le L^{(G)} (g) \le \int_{D_{\text{min}}}^{D_{\text{max}}} L(D) d(\underline{g}_i(D)).
	\label{eq: ub and lb with simple functions}
	\end{equation}
	Since $L(\cdot)$ and $g(\cdot)$ are bounded, the integral $\int_{D_{\text{min}}}^{D_{\text{max}}} L(D) d(g(D))$ exists. Therefore, in order to prove
	\begin{equation}
	L^{(G)} (g) = \int_{D_{\text{min}}}^{D_{\text{max}}} L(D) d(g(D)),
	\label{eq: convergence of the leakage to the integral}
	\end{equation}
	it suffices to show that 
	\begin{equation}
	\lim_{i \rightarrow \infty} \int_{D_{\text{min}}}^{D_{\text{max}}} L(D) d(\underline{g}_i(D)) = \int_{D_{\text{min}}}^{D_{\text{max}}} L(D) d(g(D)),
	\end{equation}
	and similarly for the integral with respect to $d(\overline{g}_i(D))$. In order to do so, we use the uniform convergence of $\underline{g}_i$ to $g$, and integration by parts. Since $L(\cdot)$ is a convex, and therefore, continuous function, the Lebesgue–--Stieltjes integral $\int_{D_{\text{min}}}^{D_{\text{max}}} L(D) d(g(D))$ reduces to a Riemann–--Stieltjes integral, and admits integration by parts \cite{Hille1996}. Thus, we can bound the difference of the two integrals as
	\begin{subequations}
		\begingroup
		\allowdisplaybreaks
		\begin{align}
		&\left|\int_{D_{\text{min}}}^{D_{\text{max}}} L(D) d(\underline{g}_i(D)) -  \int_{D_{\text{min}}}^{D_{\text{max}}} L(D) d(g(D))\right|\\
		& = \Bigg | L(D) \left( \underline{g}_i(D)\big|^{D_{\text{max}}}_{D_{\text{min}}}  - g(D)\big|^{D_{\text{max}}}_{D_{\text{min}}} \right) \nonumber \\
		& \quad + \int_{D_{\text{min}}}^{D_{\text{max}}} \left(\underline{g}_i(D)-g(D)\right) d(L(D))  \Bigg |\\
		& \le \left|L(D) \left( \underline{g}_i(D)\big|^{D_{\text{max}}}_{D_{\text{min}}}  - g(D)\big|^{D_{\text{max}}}_{D_{\text{min}}} \right)\right| \nonumber\\
		& \quad  + \left| \int_{D_{\text{min}}}^{D_{\text{max}}} \left(\underline{g}_i(D)-g(D)\right) d(L(D))  \right |,
		\end{align}
		\endgroup
	\end{subequations}
	which goes to zero as $i \rightarrow \infty$, due to uniform convergence of $\underline{g}_i$ to $g$. One can also verify the same argument for $d(\overline{g}_i(D))$. Hence, both of the integrals in \eqref{eq: ub and lb with simple functions} converge to the same value $\int_{D_{\text{min}}}^{D_{\text{max}}} L(D) d(g(D))$, and therefore \eqref{eq: convergence of the leakage to the integral} holds.

	\subsection{Proof of Lemma \ref{lemma: single letter leakage for doubly symmetric source}} \label{proof: single letter leakage for doubly symmetric source}
\begin{figure}[htb!]
		\centering
		\includegraphics[width= 0.45 \columnwidth]{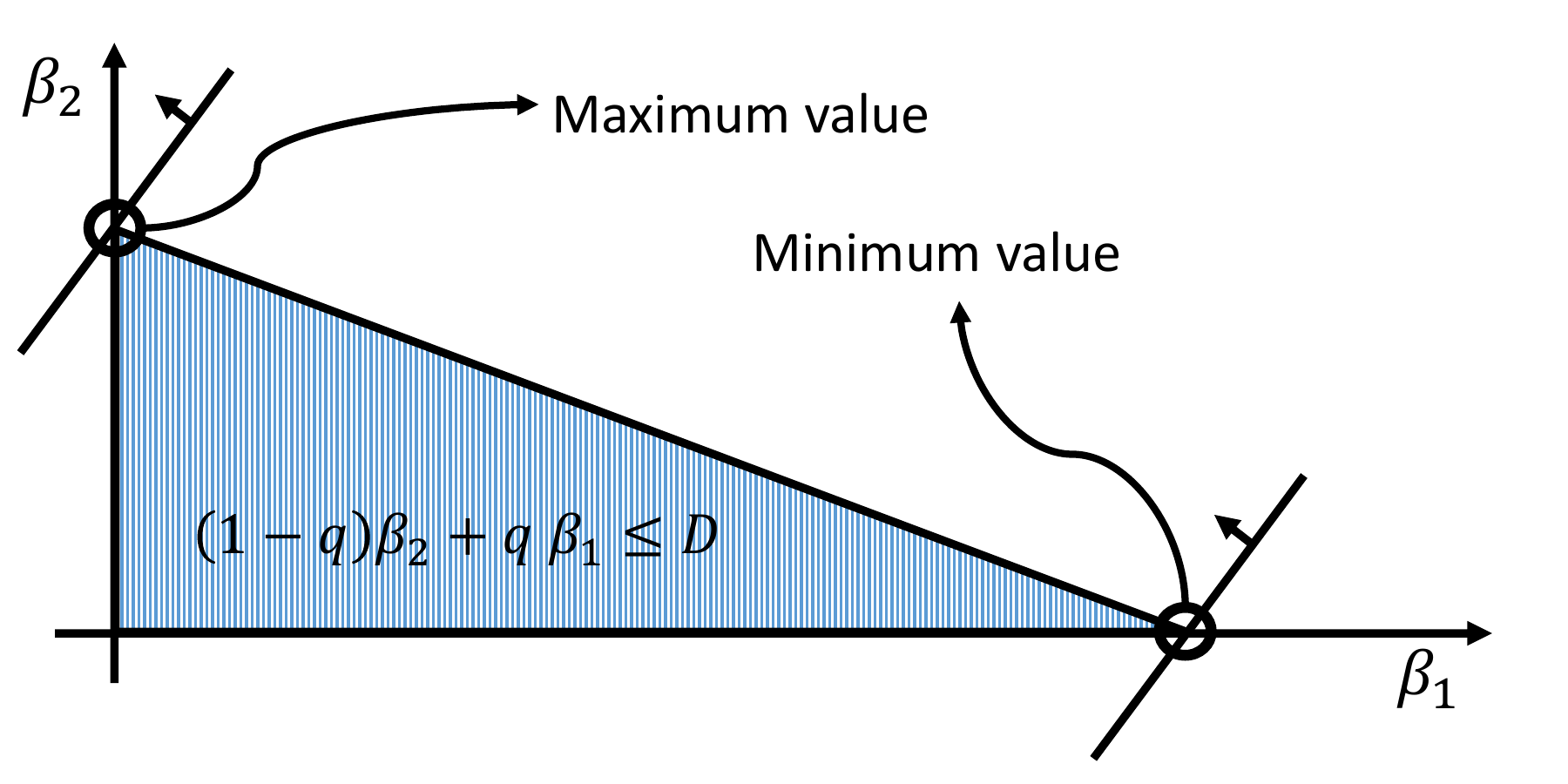}
		\caption{The feasible set and extreme values for $(1-q)\beta_2 + q(1-\beta_1)$ subject to $(1-q)\beta_2 + q\beta_1 \le D$, if $D < q$.}
		\label{fig: feasible set for LP}
	\end{figure}

	Due to the symmetry of the source distribution, and convexity of mutual information in conditional distribution, there exists an optimal mechanism with
	\begin{align}
	P(\hX=1|X=0,Y=1)&=P(\hX=0|X=1,Y=0)&=\beta_1,\\
	P(\hX=1|X=0,Y=0)&=P(\hX=0|X=1,Y=1)&= \beta_2.
	\end{align}
	Therefore, it suffices to optimize over all feasible values of $\beta_1$ and $\beta_2$. Rewriting the joint distribution $P_{Y,\hX}$ in terms of $\beta_1$, $\beta_2$, and $q$ gives
	\begin{align}
	P(Y=0,\hX=1) &= P(Y=1,\hX=0) \nonumber\\
	&= 0.5 \left[(1-q) \beta_2 + q (1-\beta_1) \right]\label{eq: Y Xhat joint 1},\\
	P(Y=0,\hX=0) &= P(Y=1,\hX=1) \nonumber\\
	&= 0.5 \left[(1-q) (1-\beta_2) + q \beta_1 \right] \label{eq: Y Xhat joint 2}.
	\end{align}
	Therefore, we have
	\begin{subequations}
		\begingroup
		\allowdisplaybreaks
		\begin{align}
		L(D) &= \min_{\substack{0 \le \beta_1,\beta_2 \le 1: \\ (1-q)\beta_2 + q\beta_1 \le D}} H(\hX) - H(\hX|Y) \\
		&= \min_{\substack{0 \le \beta_1,\beta_2 \le 1: \\ (1-q)\beta_2 + q\beta_1 \le D}} 1 - H_b\left((1-q)\beta_2 + q(1-\beta_1)\right) \label{eq: lemma: single letter leakage for doubly symmetric source: simplification 1}\\
		&=\min_{\substack{q- D \le \gamma \le q+D,\\ 0 \le \gamma \le 1}} 1 - H_b(\gamma) \label{eq: lemma: single letter leakage for doubly symmetric source: simplification 2}\\
		&= \begin{cases}
		1 - H_b(q + D), & D < 0.5 - q,\\
		0, &D\ge 0.5 - q.
		\end{cases} \label{eq: lemma: single letter leakage for doubly symmetric source: simplification 3}
		\end{align}
		\endgroup
	\end{subequations}
	where 
	\begin{itemize}
		\item \eqref{eq: lemma: single letter leakage for doubly symmetric source: simplification 1} is due to  \eqref{eq: Y Xhat joint 1} and \eqref{eq: Y Xhat joint 2},
		\item \eqref{eq: lemma: single letter leakage for doubly symmetric source: simplification 2} holds because $q \le 0.5$ and the minimum and maximum values of $(1-q)\beta_2 + q(1-\beta_1)$ subject to $(1-q)\beta_2 + q\beta_1 \le D$ are $\min\{q+D, 1\}$ and $\max\{q-D,0\}$, respectively.
		
		If $D < q$, then the extreme values occur at the corner points of the feasible region with $(\beta_1 = 0, \beta_2 = \frac{D}{1-q})$, and $(\beta_1 = \frac{D}{q}, \beta_2 = 0)$. Otherwise, if $ q \le D \le 1-q$, then the minimum and maximum values will be $0$ and $q+D$, respectively. Finally, for $D > 1-q$ the extreme values will be $0$ and $1$. The first scenario is depicted in Fig. \ref{fig: feasible set for LP}.
		\item \eqref{eq: lemma: single letter leakage for doubly symmetric source: simplification 2} is due to the fact that the binary entropy function $H_b(\cdot)$ is concave and maximized at $0.5$.
	\end{itemize}
	
	
	\section{Conclusion}
	
	We have formulated the tradeoff between privacy and utility as a minimization of mutual information between private and released data subject to two different forms of distortion constraints: the average distortion cost constraint and the complementary CDF bound on distortion. The former allows for taking non-separable distortion measures into account, while the latter enables the data publisher to provide refined guarantees on utility.
	
	For the average distortion cost constraints, we have characterized the asymptotically optimal leakage for both stationary memoryless and general mechanisms as a function of the single letter leakage function $L$ and the distortion cost function $f$. In particular, we have shown that a memoryless mechanism achieves the asymptotically optimal leakage if and only if the information leakage-cost function $L(f^{-1}(\cdot))$ coincides with its lower convex envelope; otherwise, a mixture of exactly two memoryless mechanisms is sufficient.
	
	For the complementary CDF bound on distortion, we have derived the asymptotically optimal leakage. We have shown that under general mechanisms the optimal leakage is equal to the integral of the single letter leakage function with respect to the Lebesgue---Stieltjes measure associated with the complementary CDF bound, while for stationary and memoryless mechanisms, it is equal to the single letter leakage function evaluated at the largest value of distortion for which the CDF bound function is equal to one.

	For both types of utility constraints, the challenge remains to characterize the second order performance of the leakage as a function of the data size $n$. More generally, the proof techniques developed here for arbitrary cost functions and complementary CDF bounds on distortion are applicable to a broad class of information theoretic problems such as lossy source coding with fidelity constraints and channel coding with input cost constraints.
	

	\section*{Disclosure of AI Assistance}

	OpenAI Codex was used in preparing this updated version of the manuscript to facilitate \LaTeX{} typesetting and to support exploratory discussion of proof ideas for the revised mathematical arguments. The authors reviewed and verified the final text, proofs, and citations, and remain responsible for the content of the manuscript.
	
	
	\bibliographystyle{IEEEtran}
	
	\bibliography{IEEEabrv,Privacy}

@InProceedings{CalmonMM15,
  Title                    = {Fundamental limits of perfect privacy},
  Author                   = {Fl{\'{a}}vio du Pin Calmon and Ali Makhdoumi and Muriel M{\'{e}}dard},
  Booktitle                = {{IEEE} International Symposium on Information Theory, {ISIT} 2015, Hong Kong, China, June 14-19, 2015},
  Year                     = {2015},
  Pages                    = {1796--1800},
  Bibsource                = {dblp computer science bibliography, http://dblp.org},
  DOI                      = {10.1109/ISIT.2015.7282765},
  URL                      = {http://dx.doi.org/10.1109/ISIT.2015.7282765}
}

@Article{Asoodeh2015,
  Title                    = {Information Extraction Under Privacy Constraints},
  Author                   = {Shahab Asoodeh and Mario Diaz and Fady Alajaji and Tam{\'{a}}s Linder},
  Journal                  = {CoRR},
  Year                     = {2015},
  Volume                   = {abs/1511.02381},
  Bibsource                = {dblp computer science bibliography, http://dblp.org}
}

@InProceedings{CalmonFawaz2013,
  Title                    = {Privacy against statistical inference},
  Author                   = {F. du Pin Calmon and N. Fawaz},
  Booktitle                = {2012 50th Annual Allerton Conference on Communication, Control, and Computing (Allerton)},
  Year                     = {2012},
  Month                    = {Oct},
  Pages                    = {1401-1408},
  DOI                      = {10.1109/Allerton.2012.6483382}
}

@InProceedings{Duchi,
  Title                    = {Local Privacy and Statistical Minimax Rates},
  Author                   = {Duchi, J.C. and Jordan, M.I. and Wainwright, M.J.},
  Booktitle                = {Foundations of Computer Science (FOCS), 2013 IEEE 54th Annual Symposium on},
  Year                     = {2013},
  Month                    = {Oct},
  Pages                    = {429-438},
  DOI                      = {10.1109/FOCS.2013.53},
  ISSN                     = {0272-5428}
}

@Book{Hille1996,
  Title                    = {Functional analysis and semi-groups},
  Author                   = {Hille, Einar and Phillips, Ralph Saul},
  Publisher                = {American Mathematical Society},
  Year                     = {1996},
  Volume                   = {31}
}

@Article{Hoeffding,
  Title                    = {Probability Inequalities for Sums of Bounded Random Variables},
  Author                   = {Wassily Hoeffding},
  Journal                  = {Journal of the American Statistical Association},
  Year                     = {1963},
  Number                   = {301},
  Pages                    = {13-30},
  Volume                   = {58},
  ISSN                     = {01621459},
  Publisher                = {[American Statistical Association, Taylor \& Francis, Ltd.]},
  URL                      = {http://www.jstor.org/stable/2282952}
}

@Article{Kairouz,
  Title                    = {Extremal Mechanisms for Local Differential Privacy},
  Author                   = {Kairouz, Peter and Oh, Sewoong and Viswanath, Pramod},
  Journal                  = {J. Mach. Learn. Res.},
  Year                     = {2016},
  Month                    = jan,
  Number                   = {1},
  Pages                    = {492--542},
  Volume                   = {17},
  Acmid                    = {2946662},
  ISSN                     = {1532-4435},
  Issue_date               = {January 2016},
  Numpages                 = {51},
  Publisher                = {JMLR.org},
  URL                      = {http://dl.acm.org/citation.cfm?id=2946645.2946662}
}

@InProceedings{KKSAllerton2016,
  Title                    = {On the Fine Asymptotics of Information Theoretic Privacy},
  Author                   = {K. Kalantari and O. Kosut and L. Sankar},
  Booktitle                = {2016 54th Annual Allerton Conference on Communication, Control, and Computing (Allerton)},
  Year                     = {2016}
}

@Article{KLNRS11,
  Title                    = {What Can We Learn Privately?},
  Author                   = {Shiva Prasad Kasiviswanathan and Homin K. Lee and Kobbi Nissim and Sofya Raskhodnikova and Adam Smith},
  Journal                  = {SIAM Journal on Computing},
  Year                     = {2008},
  Number                   = {3},
  Pages                    = {793--826},
  Volume                   = {40},
  Booktitle                = {SIAM Journal on Computing},
  DOI                      = {10.1137/090756090}
}

@Article{Kostina,
  Title                    = {Variable-Length Compression Allowing Errors},
  Author                   = {V. Kostina and Y. Polyanskiy and S. Verd{\'u}},
  Journal                  = {IEEE Transactions on Information Theory},
  Year                     = {2015},
  Month                    = {Aug},
  Number                   = {8},
  Pages                    = {4316-4330},
  Volume                   = {61},
  DOI                      = {10.1109/TIT.2015.2438831},
  ISSN                     = {0018-9448}
}

@Article{Kostina2012c,
  Title                    = {Fixed-Length Lossy Compression in the Finite Blocklength Regime},
  Author                   = {V. Kostina and S. Verd{\'u}},
  Journal                  = {IEEE Trans. Inform. Theory},
  Year                     = {2012},
  Month                    = {Jun.},
  Number                   = {6},
  Pages                    = {3309-3338},
  Volume                   = {58}
}

@InProceedings{KoushaISIT2016,
  Title                    = {Optimal differential privacy mechanisms under Hamming distortion for structured source classes},
  Author                   = {K. Kalantari and L. Sankar and A. D. Sarwate},
  Booktitle                = {2016 IEEE International Symposium on Information Theory (ISIT)},
  Year                     = {2016},
  Month                    = {July},
  Pages                    = {2069-2073},
  DOI                      = {10.1109/ISIT.2016.7541663}
}

@InProceedings{KoushaISIT2017,
  Title                    = {On information-theoretic privacy with general distortion cost functions},
  Author                   = {K. Kalantari and L. Sankar and O. Kosut},
  Booktitle                = {2017 IEEE International Symposium on Information Theory (ISIT)},
  Year                     = {2017},
  Month                    = {June},
  Pages                    = {2865-2869},
  DOI                      = {10.1109/ISIT.2017.8007053}
}

@Article{Sankar_TIFS_2013,
  Title                    = {Utility-Privacy Tradeoffs in Databases: An Information-Theoretic Approach},
  Author                   = {L. Sankar and S. R. Rajagopalan and H. V. Poor},
  Journal                  = {IEEE Transactions on Information Forensics and Security},
  Year                     = {2013},
  Number                   = {6},
  Pages                    = {838--852},
  Volume                   = {8},
  DOI                      = {10.1109/TIFS.2013.2253320}
}

@InProceedings{Shkel,
  Title                    = {A coding theorem for f-separable distortion measures},
  Author                   = {Y. Shkel and S. Verd{\'u}},
  Booktitle                = {2016 Information Theory and Applications Workshop (ITA)},
  Year                     = {2016},
  Month                    = {Jan},
  Pages                    = {1-7},
  DOI                      = {10.1109/ITA.2016.7888172}
}

@Article{tan2014asymptotic,
  Title                    = {Asymptotic Estimates in Information Theory with Non-Vanishing Error Probabilities},
  Author                   = {Vincent Y. F. Tan},
  Journal                  = {Foundations and Trends® in Communications and Information Theory},
  Year                     = {2014},
  Number                   = {1-2},
  Pages                    = {1-184},
  Volume                   = {11},
  DOI                      = {10.1561/0100000086},
  ISSN                     = {1567-2190}
}

@Article{Warner,
  Title                    = {Randomized Response: A Survey Technique for Eliminating Evasive Answer Bias},
  Author                   = {Stanley L. Warner},
  Journal                  = {Journal of the American Statistical Association},
  Year                     = {1965},
  Number                   = {309},
  Pages                    = {63-69},
  Volume                   = {60},
  ISSN                     = {01621459},
  Publisher                = {[American Statistical Association, Taylor \& Francis, Ltd.]}
}

@Article{Yamamoto,
  Title                    = {A source coding problem for sources with additional outputs to keep secret from the receiver or wiretappers (Corresp.)},
  Author                   = {H. Yamamoto},
  Journal                  = {IEEE Transactions on Information Theory},
  Year                     = {1983},
  Month                    = {Nov},
  Number                   = {6},
  Pages                    = {918-923},
  Volume                   = {29},
  DOI                      = {10.1109/TIT.1983.1056749},
  ISSN                     = {0018-9448}
}

@Article{YannisVerdu,
  Title                    = {Optimal Lossless Data Compression: Non-Asymptotics and Asymptotics},
  Author                   = {Kontoyiannis, I. and Verd{\'u}, S.},
  Journal                  = {Information Theory, IEEE Transactions on},
  Year                     = {2014},
  Month                    = {Feb},
  Number                   = {2},
  Pages                    = {777-795},
  Volume                   = {60},
  DOI                      = {10.1109/TIT.2013.2291007},
  ISSN                     = {0018-9448}
}

@Book{Yeung,
  Title                    = {Information Theory and Network Coding},
  Author                   = {Yeung, Raymond W.},
  Publisher                = {Springer Publishing Company, Incorporated},
  Year                     = {2008},
  Edition                  = {1},
  ISBN                     = {0387792333, 9780387792330}
}

@InProceedings{YucelAllerton2015,
  Title                    = {Variable-length channel codes with probabilistic delay guarantees},
  Author                   = {Y. Altu\u{g} and H. V. Poor and S. Verd{\'u}},
  Booktitle                = {2015 53rd Annual Allerton Conference on Communication, Control, and Computing (Allerton)},
  Year                     = {2015},
  Month                    = {Sept},
  Pages                    = {642-649},
  DOI                      = {10.1109/ALLERTON.2015.7447065}
}

@InProceedings{YucelISIT2015,
  Title                    = {On fixed-length channel coding with feedback in the moderate deviations regime},
  Author                   = {Y. Altu\u{g} and H. V. Poor and S. Verd{\'u}},
  Booktitle                = {2015 IEEE International Symposium on Information Theory (ISIT)},
  Year                     = {2015},
  Month                    = {June},
  Pages                    = {1816-1820},
  DOI                      = {10.1109/ISIT.2015.7282769},
  ISSN                     = {2157-8095}
}

@InProceedings{YucelISIT2016,
  Title                    = {On channel dispersion per unit cost},
  Author                   = {Y. Altu\u{g} and H. V. Poor and S. Verd{\'u}},
  Booktitle                = {2016 IEEE International Symposium on Information Theory (ISIT)},
  Year                     = {2016},
  Month                    = {July},
  Pages                    = {2429-2433},
  DOI                      = {10.1109/ISIT.2016.7541735}
}

@InProceedings{Calmon2013,
  author    = {F. P. Calmon and M. Varia and M. Médard and M. M. Christiansen and K. R. Duffy and S. Tessaro},
  title     = {Bounds on inference},
  booktitle = {2013 51st Annual Allerton Conference on Communication, Control, and Computing (Allerton)},
  year      = {2013},
  pages     = {567-574},
  month     = {Oct},
}

@Article{AsoodehDAL2019,
  author  = {Shahab Asoodeh and Mario Diaz and Fady Alajaji and Tam{\'{a}}s Linder},
  journal = {IEEE Transactions on Information Theory},
  title   = {Estimation Efficiency Under Privacy Constraints},
  year    = {2019},
  number  = {3},
  pages   = {1512--1534},
  volume  = {65},
  doi     = {10.1109/TIT.2018.2865558},
}

@Article{WangVCMDV2019,
  author  = {Hao Wang and Lisa Vo and Flavio P. Calmon and Muriel M{\'{e}}dard and Ken R. Duffy and Mayank Varia},
  journal = {IEEE Transactions on Information Theory},
  title   = {Privacy With Estimation Guarantees},
  year    = {2019},
  number  = {12},
  pages   = {8025--8042},
  volume  = {65},
  doi     = {10.1109/TIT.2019.2934414},
}

@Article{LiaoTIT2019,
  author  = {Jiachun Liao and Oliver Kosut and Lalitha Sankar and Flavio P. Calmon},
  journal = {IEEE Transactions on Information Theory},
  title   = {Tunable Measures for Information Leakage and Applications to Privacy-Utility Tradeoffs},
  year    = {2019},
  number  = {12},
  pages   = {8043--8066},
  volume  = {65},
  doi     = {10.1109/TIT.2019.2935768},
}

@Article{IssaTIT2020,
  author  = {Ibrahim Issa and Aaron B. Wagner and Sudeep Kamath},
  journal = {IEEE Transactions on Information Theory},
  title   = {An Operational Approach to Information Leakage},
  year    = {2020},
  number  = {3},
  pages   = {1625--1657},
  volume  = {66},
  doi     = {10.1109/TIT.2019.2962804},
}

@Article{DiazWCS2020,
  author  = {Mario Diaz and Hao Wang and Flavio P. Calmon and Lalitha Sankar},
  journal = {IEEE Transactions on Information Theory},
  title   = {On the Robustness of Information-Theoretic Privacy Measures and Mechanisms},
  year    = {2020},
  number  = {4},
  pages   = {1949--1978},
  volume  = {66},
  doi     = {10.1109/TIT.2019.2939472},
}

@Article{RassouliGunduz2020TV,
  author  = {Borzoo Rassouli and Deniz G{\"{u}}nd{\"{u}}z},
  journal = {IEEE Transactions on Information Forensics and Security},
  title   = {Optimal Utility-Privacy Trade-Off With Total Variation Distance as a Privacy Measure},
  year    = {2020},
  pages   = {594--603},
  volume  = {15},
  doi     = {10.1109/TIFS.2019.2903658},
}

@Article{RassouliGunduz2021Perfect,
  author  = {Borzoo Rassouli and Deniz G{\"{u}}nd{\"{u}}z},
  journal = {IEEE Journal on Selected Areas in Information Theory},
  title   = {On Perfect Privacy},
  year    = {2021},
  number  = {1},
  pages   = {177--191},
  volume  = {2},
  doi     = {10.1109/JSAIT.2021.3053432},
}

@Article{ZamaniOS2021,
  author  = {Amirreza Zamani and Tobias J. Oechtering and Mikael Skoglund},
  journal = {IEEE Transactions on Information Forensics and Security},
  title   = {A Design Framework for Strongly $\chi^2$-Private Data Disclosure},
  year    = {2021},
  pages   = {2312--2325},
  volume  = {16},
  doi     = {10.1109/TIFS.2021.3053462},
}

@Article{ZamaniOS2022,
  author  = {Amirreza Zamani and Tobias J. Oechtering and Mikael Skoglund},
  journal = {IEEE Transactions on Information Forensics and Security},
  title   = {Data Disclosure With Non-Zero Leakage and Non-Invertible Leakage Matrix},
  year    = {2022},
  pages   = {165--179},
  volume  = {17},
  doi     = {10.1109/TIFS.2021.3137755},
}

@Article{ZamaniOS2024,
  author  = {Amirreza Zamani and Tobias J. Oechtering and Mikael Skoglund},
  journal = {IEEE Transactions on Information Theory},
  title   = {On the Privacy-Utility Trade-Off With and Without Direct Access to the Private Data},
  year    = {2024},
  number  = {3},
  pages   = {2177--2200},
  volume  = {70},
  doi     = {10.1109/TIT.2023.3326070},
}

@Article{KurriSK2024,
  author  = {Gowtham R. Kurri and Lalitha Sankar and Oliver Kosut},
  journal = {IEEE Transactions on Information Theory},
  title   = {An Operational Approach to Information Leakage via Generalized Gain Functions},
  year    = {2024},
  number  = {2},
  pages   = {1349--1375},
  volume  = {70},
  doi     = {10.1109/TIT.2023.3341148},
}

@Article{GilaniKKS2024,
  author  = {Atefeh Gilani and Gowtham R. Kurri and Oliver Kosut and Lalitha Sankar},
  journal = {IEEE Transactions on Information Theory},
  title   = {Unifying Privacy Measures via Maximal $(\alpha,\beta)$-Leakage},
  year    = {2024},
  number  = {6},
  pages   = {4368--4395},
  volume  = {70},
  doi     = {10.1109/TIT.2024.3384922},
}

@Article{SaeidianGSSO2026,
  author  = {Sara Saeidian and Leonhard Grosse and Parastoo Sadeghi and Mikael Skoglund and Tobias J. Oechtering},
  journal = {IEEE Transactions on Information Theory},
  title   = {Information Density Bounds for Privacy},
  year    = {2026},
  number  = {1},
  pages   = {610--635},
  volume  = {72},
  doi     = {10.1109/TIT.2025.3637364},
}

@Article{TaylorVC2026,
  author  = {S. Taylor and P. K. Vippathalla and J. P. Coon},
  journal = {IEEE Transactions on Information Theory},
  title   = {The Asymptotic Behavior of Information Leakage Metrics},
  year    = {2026},
  number  = {2},
  pages   = {811--831},
  volume  = {72},
  doi     = {10.1109/TIT.2025.3646586},
}

@InProceedings{LiaoKSC2018Hard,
  author    = {Jiachun Liao and Oliver Kosut and Lalitha Sankar and Flavio P. Calmon},
  booktitle = {2018 IEEE Information Theory Workshop (ITW)},
  title     = {Privacy Under Hard Distortion Constraints},
  year      = {2018},
  pages     = {1--5},
  doi       = {10.1109/ITW.2018.8613385},
}

@InProceedings{SreekumarGunduz2019,
  author    = {Sreejith Sreekumar and Deniz G{\"{u}}nd{\"{u}}z},
  booktitle = {2019 IEEE International Symposium on Information Theory (ISIT)},
  title     = {Optimal Privacy-Utility Trade-off Under a Rate Constraint},
  year      = {2019},
  pages     = {2159--2163},
  doi       = {10.1109/ISIT.2019.8849330},
}

@InProceedings{ZhaoCTG2020,
  author    = {Han Zhao and Jianfeng Chi and Yuan Tian and Geoffrey J. Gordon},
  booktitle = {Advances in Neural Information Processing Systems},
  title     = {Trade-offs and Guarantees of Adversarial Representation Learning for Information Obfuscation},
  year      = {2020},
  volume    = {33},
}

@InProceedings{GuoSS2023,
  author    = {Chuan Guo and Alexandre Sablayrolles and Maziar Sanjabi},
  booktitle = {Proceedings of the 40th International Conference on Machine Learning},
  title     = {Analyzing Privacy Leakage in Machine Learning via Multiple Hypothesis Testing: A Lesson From {Fano}},
  year      = {2023},
  pages     = {11998--12011},
  publisher = {PMLR},
  series    = {Proceedings of Machine Learning Research},
  volume    = {202},
  url       = {https://proceedings.mlr.press/v202/guo23e.html},
}
	
	%
	
	
	
\end{document}